\documentclass[aps,twocolumn,prd,superscriptaddress,showpacs,nofootinbib]{revtex4-1}

\usepackage[utf8]{inputenc}
\usepackage{diagbox}

\usepackage{mathtools}
\usepackage{amsfonts}
\usepackage{mathrsfs}
\usepackage{bbm}
\usepackage{slashed}

\usepackage{graphicx}
\usepackage{color}
\usepackage{array}

\usepackage{placeins}
\usepackage{booktabs}
\usepackage{makecell}
\usepackage{epstopdf}
\usepackage[caption=false]{subfig}
\usepackage{multirow}

\usepackage{xspace}
\usepackage{siunitx}
\usepackage{hyperref}
\usepackage[nameinlink]{cleveref}
\usepackage{appendix}

\usepackage{xifthen}
\usepackage{xcolor}
\hypersetup{
	colorlinks,
	linkcolor={red!75!black},
	citecolor={blue!75!black},
	urlcolor={blue!75!black}
}

\newcommand{\gettitle}{Extracting Hadron-Quark Phase Transition Chemical Potential via Astronomical Observations}
\hypersetup{
	pdftitle={\gettitle},
	pdfauthor={Bai, Liu},
	pdfkeywords={neutron star}
		{tidal deformability} {phase structure}
	 {Dyson-Schwinger equation},
	bookmarksopen=true,
	bookmarksopenlevel=2,
	bookmarksnumbered=true
}

\begin{document}

\title{\gettitle}
\author{Zhan Bai}
\email{{baizhan@pku.edu.cn.\ Present address: Institute of Theoretical } 
\\
{ Physics, Chinese academy of Science, Beijing 100081, China.} }
\affiliation{Department of Physics and State Key Laboratory of Nuclear Physics and Technology, Peking University, Beijing 100871, China}
\affiliation{Collaborative Innovation Center of Quantum Matter, Beijing 100871, China}

\author{Yuxin Liu}
\email[Corresponding author: ]{yxliu@pku.edu.cn}
\affiliation{Department of Physics and State Key Laboratory of Nuclear Physics and Technology, Peking University, Beijing 100871, China}
\affiliation{Collaborative Innovation Center of Quantum Matter, Beijing 100871, China}
\affiliation{Center for High Energy Physics, Peking University, Beijing 100871, China}

%\pacs{25.75.Nq, %Quark deconfinement, quark-gluon plasma production, and phase transitions
%%      04.30.Tv, %Gravitational-wave astrophysics
%      26.60.Kp, %Equations of state of neutron-star matter
%     }                             % Classification Scheme.

\begin{abstract}
We propose a scheme to identify the region for the hadron-quark phase transition in cold dense strong interaction matter (neutron star matter) to take place with astronomical observations.
To study the property of the first-order phase transition in the matter,
we take the sound speed as the interpolation objective to construct the equation of state of the matter involving the hadron and quark coexisting phase.
We show the feature of the sound speed in the matter with two conservation charges and a first--order phase transition.
With the maximum mass, the tidal deformability and the radius of neutron stars being taken as calibration quantities,
the baryon chemical potential region for the phase transition to occur is constrained to a narrow range,
and the most probable value of the phase transition chemical potential is found.
\end{abstract}

\maketitle

\section{\label{sec:Intro}Introduction}
Researches on the phase structure of strong interaction matter have attracted great attentions for decades.
It is generally believed that at low temperature and density,
the basic degree of freedom of strong interaction matter is the color-singlets, i.e. hadrons,
while at high temperature and/or high density,
the quarks deconfine from hadron and become the basic degree of freedom.
The phase transition from hadron to quark matter is usually referred to as hadron-quark phase transition,
or simply QCD phase transition.

At low density and high temperature, the QCD phase transition has been understood well both theoretically and experimentally.
On theoretical side, the lattice QCD, the Dyson-Schwinger (DS) equation approach, the functional renormalization group (FRG) approach
and many effective models have made great progress which shows that the phase evolution in case of the physical quark mass is in fact a crossover but not finite-order phase transition
(see, e.g., Refs.~\cite{Ding:2015ona,Roberts:2000aa,Qin:2010nq,Bashir:2012fs,Gao:2016qkh,Fischer:2018sdj,Herbst:2013ail,Fister:2013bh,Fu:2019hdw,Gao:2020qsj,Fukushima:2017csk}).
On experimental side, one can generate such kind of matter with relativistic heavy ion collisions (RHICs) in laboratory.

However, for the QCD matter in low temperature and high density (i.e. cold dense matter),
the theoretical situation is more severe,
since the first-principle lattice QCD simulation is not adaptable because of the notorious sign problem,
and reliable experimental data are required to calibrate the calculations in the framework of the DS equation approach and the FRG approach.
The density and chemical potential for the phase transition to take place is then still not known.
There are even debates on whether there exists a phase transition.
Although DS equation, FRG and effective model calculations indicate that there should be a first-order phase transition in the region~\cite{Qin:2010nq,Gao:2016qkh,Fischer:2018sdj,Fu:2019hdw,Gao:2020qsj,Fu:2007xc,Fukushima:2008wg,Fukushima:2011jc,Gao:2015kea},
there are also arguments that the transition should be a smooth crossover,
i.e., no phase transition at all~\cite{Brandes:2021pti,Baym:2017whm,Fukushima:2020cmk}.
And on experimental side, the corresponding density is far beyond the capability of current terrestrial experiments.

A complementary method beside the RHIC experiments is the observation of dense astronomical objects,
especially those for compact stars.
Neutron star is one of the most compact object in the universe,
and it is very likely that the density inside neutron star is so large that there exists a hadron-quark phase transition~\cite{Glendenning:2000,Baym:2017whm,Oertel:2016bki,Lattimer:2004pg,Weber:2006ep}.
When the phase transition occurs inside neutron stars,
the equation of state (EoS) of the matter will be changed, and so do the mass-radius relation and the maximum mass of the stars.
There have already been several neutron stars with large mass observed~\cite{Demorest:2010bx,Antoniadis:2013pzd,Fonseca:2016tux,NANOGrav:2017wvv,Linares:2018ppq,NANOGrav:2019jur,Fonseca:2021wxt,Riley:2021pdl,Miller:2021qha},
	which requires that the EoS should be stiff.
The gravitational wave from binary neutron star merger also provides important information about the EoS~\cite{LIGOScientific:2017vwq,LIGOScientific:2017zic,LIGOScientific:2020aai}.
From the gravitational wave, it is found that the tidal deformability of the star should be small, and hence the EoS should be rather soft.
Also, recent NICER provides significant information for the profile of the neutron star~\cite{Riley:2019yda,Miller:2019cac,Riley:2021pdl,Miller:2021qha}.
These indicate that astronomical observations have provided strict constraints on the EoS.
For example, in Refs.~\cite{Annala:2017llu,Annala:2019puf},
the EoS from $n_{B}=1.1n_{s}$ to $\mu_{B}\le 2.6$GeV have been constrained to a rather narrow range,
where $n_{s}$ is the nuclear saturation density.
We then take these recent observation data to identify the phase transition and the chemical potential (density) region for the phase transition to occur.

Because of the lack of fundamental approach to provide the EoS of the matter involving the hadron-quark phase transitions,
one usually take the way that describes the hadron matter and the quark matter separately via respective approach,
and combine them together to get the complete EoS in the whole density region (with hadron matter, hybrid matter, quark matter, seperately).
It has been known that the most widely used construction schemes are the Maxwell construction and the Gibbs construction~\cite{Glendenning:2000}.
Both of these schemes describe a first-order phase transition.
The Maxwell construction gives a violent phase transition with a density gap,
i.e., the EoS is not continuous.
The Gibbs construction, on the other hand, gives first-order phase transition with two conservation charges,
and the phase transition is milder.
Under Gibbs construction, the EoS is continuous but not smooth.

In order to describe a crossover, the 3-window construction was proposed~\cite{Masuda:2012ed,Masuda:2012kf,Baym:2017whm,Kojo:2014rca,Kojo:2015fua},
and the corresponding EoS is expected to be continuous and smooth.
%{
The 3-window construction can also help reduce the uncertainties of hadron and quark models in phase transition region,
as will be discussed in detail in Sec.~\ref{sec:OldConstruction}.
%}

Apart from those constructions with models,
there have been schemes which directly construct EoS without concerning the underlying model,
for example the segmented polytropic EoS~\cite{Annala:2017llu}.

It has also been known that the sound speed (SS) is closely related to the EoS,
and has been used in the study neutron stars.
For example, the phase transition signal in neutron star oscillations is studied with SS~\cite{Wei:2018tts,Jaikumar:2021jbw,Bai:2021wrh},
and neutron star measurements have been taken to provide insights of the SS in intermediate density~\cite{Tews:2018kmu}.
The SS can then serve as a significant criterion for the phase transition.
For example, under Maxwell construction, there will be a density range with vanishing SS,
and there will be discontinuities in SS under Gibbs construction.
Under 3-window construction, however, the SS should be continuous and smooth everywhere.

There have already been studies which construct the SS in cold dense matter.
In Refs.~\cite{Alford:2013aca,Han:2018mtj,Li:2021crp},
the authors used a ``constant speed of sound construction'' which assumes that the SS for quark matter is a constant,
and the phase transition is descried with a density gap of the EoS.
This corresponds to a violent first-order phase transition similar to the Maxwell construction.
In Refs.~\cite{Greif:2018njt,Brandes:2022nxa},
the SS is constructed using a smooth Gaussian function, which indicates a crossover,
similar to the 3-window construction.
The SS can also be constructed using piecewise linear function,
which includes the most possibilities,
as has been done in Refs.~\cite{Annala:2019puf,Tews:2019cap}.
In this way, the SS is continuous but not smooth,
which means that the phase transition is neither first-order nor crossover.
However, if the number of segmentations is large enough,
this construction can approach to crossover.
Also, as argued in Ref.~\cite{Annala:2019puf},
this construction scheme can mimic the Maxwell construction closely enough.

However, in the above-mentioned progress on the SS construction,
there is no attempt to mimic the phase transition with two conservation charges,
i.e., the phase transition described by Gibbs construction.
For Gibbs construction,
there should be two discontinuities in the SS as a function of density,
and SS is not vanishing between these two discontinuities.
Some examples of such SS can be found in Fig.4 of Ref.~\cite{Han:2019bub} and Fig.6 of Ref.~\cite{Jaikumar:2021jbw}.

We then in this paper
develop a scheme to construct the SS and the EoS of the matter involving hadron-quark phase transition and two conservation charges,
and make use of them to calculate the properties of compact stars.
The construction takes the baryon chemical potential (density) as the parameter to label the phase transition.
In this way, the uncertainties of hadron and quark approaches in phase transition region can be absorbed into the uncertainties of the phase transition chemical potential.
%}
With the mass, the radius and the tidal deformability of the stars being taken as calibrations
to exclude the inappropriate parameters,
we extract then the most probable baryon chemical potential region for the hadron-quark phase transition
to happen in the cold dense matter with two conservation charges.

The paper is organized as follows.
In Sec.~\ref{sec:construction}, we describe our new method to construct the EoS of the cold compact star matter involving the hadron-quark phase transition,
where in Sec.~\ref{sec:1stOrderPT}, we introduce the basic feature of the phase transitions with two conservation charges,
and talk about the necessity to develop a new construction scheme in Sec.~\ref{sec:OldConstruction}.
The detail of our new construction scheme is described in Sec.~\ref{sec:SoundSpeedConstruction}.
In Sec.~\ref{sec:numerical}, we implement our proposed construction to calculate the property of compact star,
and take the astronomical observations to constrain the EoS of the matter as well as the possible range for the phase transition to take place.
In Sec.~\ref{sec:sum}, we give a summary and remark.

\section{Constructing the Hadron-Quark Phase Transition in Hybrid Star Matter}\label{sec:construction}

\subsection{First-Order Phase Transition with Two Conservation Charges}\label{sec:1stOrderPT}

In hybrid star matter, there are two conservation charges,
namely the baryon number and the electric charge number.
Therefore, we need two chemical potentials to identify the state of the matter,
which are usually referred to as the baryon chemical potential $\mu_{B}^{}$ and the charge chemical potential $\mu_{Q}^{}$.
In hadron phase, we have $\mu_{B}^{}=\mu_{n}^{}$ and $\mu_{Q}^{}=-\mu_{e}^{}$,
where $\mu_{n}^{}$, $\mu_{e}^{}$ is the chemical potential for neutron, electron, respectively.
In quark phase, we have $\mu_{B}^{}=\mu_{u}^{} + 2\mu_{d}^{}$,
where $\mu_{u}^{}$, $\mu_{d}^{}$ is the chemical potential of $u$, $d$ quark, respectively.

\begin{figure}[!htb]
\includegraphics[width=0.45\textwidth]{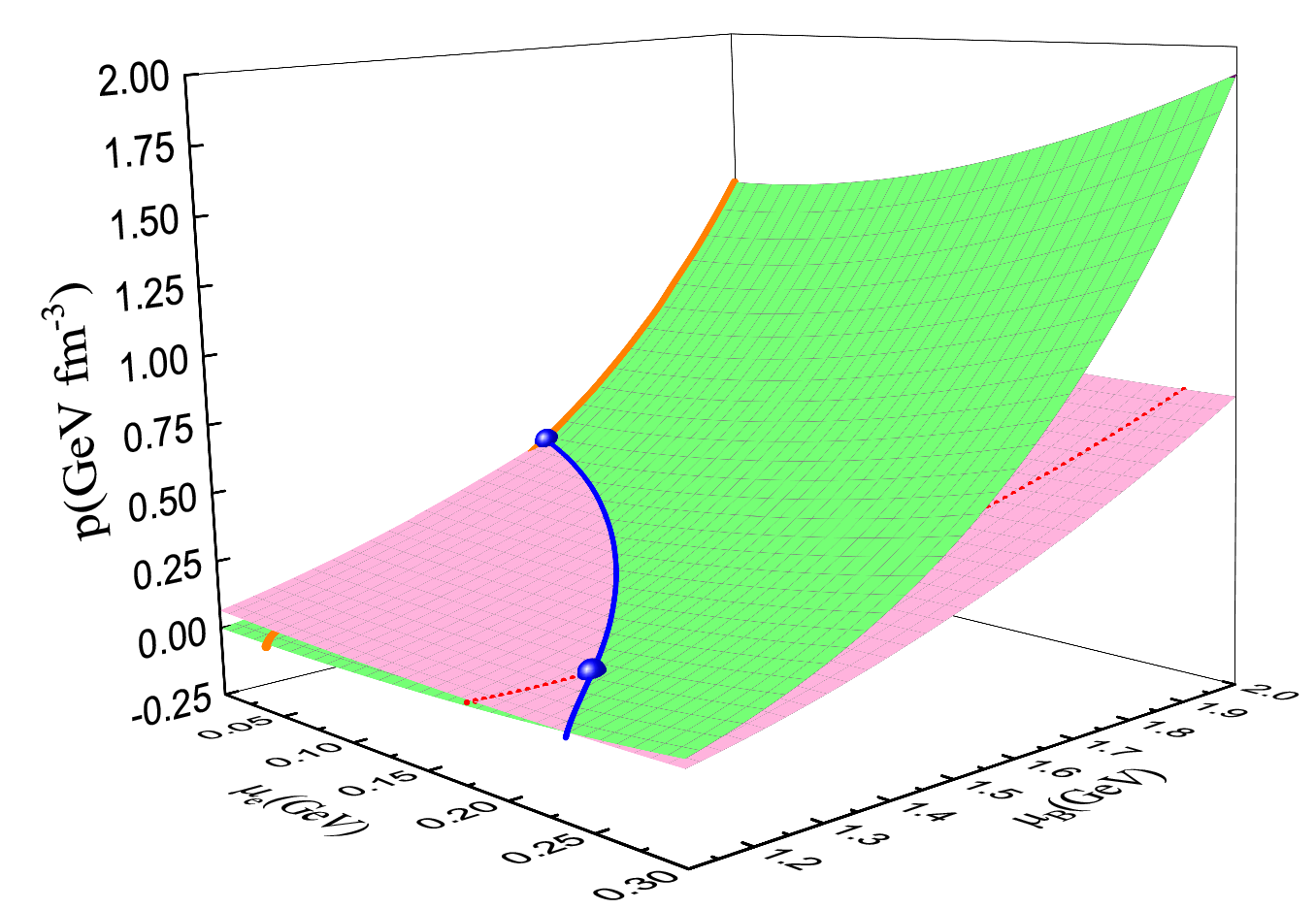}
\caption{(color online)
Calculated function surface of the pressure in terms of the two chemical potentials
of the hadron phase without including hyperons (pink) and that of the quark phase (green).
}
\label{fig:HadronQuarkSurface}
\end{figure}

Since the pressure of the matter is governed by two chemical potentials,
it appears as a surface in $p\,$--$\mu_{B}^{}$--$\mu_{e}^{}$ space, for both the hadron phase and the quark phase.
It has been known that the relativistic mean field model (RMF)~~\cite{Walecka:1974AP,Serot:1986ANP} is a successful
theory in describing the properties of hadron matter in the density region relevant in compact stars~\cite{Glendenning:2000,Oertel:2016bki}.
And the Dyson-Schwinger (DS) equation approach, almost uniquely, the scheme which includes both confinement and
chiral symmetry breaking~\cite{McLerran:2007qj},
is successful in describing hadron properties and QCD phase transitions (see, e.g., Refs.~\cite{Roberts:1994dr,Roberts:2000aa,Qin:2010nq,Bashir:2012fs,Gao:2016qkh,Maris:1997tm,Chang:2009zb,Eichmann:2009qa,Fischer:2018sdj}, 
especially for the high density (baryon chemical potential) region where the lattice QCD simulation does not work well due to the so-called ``sign problem''~\cite{Ukawa:2015}.
%}
Details of the description of the hadron and the quark approaches are represented in the appendices.
We then take the RMF with TW-99 parameterization~\cite{Typel:1999yq} (without the inclusion of hyperons) to calculate the hadron surface.
and implement the DS equation approach of QCD described in Refs.~\cite{Chen:2011my,Bai:2017wvk} with $\alpha=2$ to describe the quark surface.
The obtained results are shown in Fig.~\ref{fig:HadronQuarkSurface} as an illustration of the hadron and the quark surfaces.

It is well known that, as the phase transition happens,
the Gibbs condition should be satisfied,
{\it i.e.} the chemical potentials ($\mu_{B}^{}$, $\mu_{e}^{}$) and the pressure ($p$)
should be the same in the two phases.
Therefore, the phase transition should only happen on the cross line of the two surfaces,
which corresponds to the blue solid line shown in Fig.~\ref{fig:HadronQuarkSurface}.
At low and high baryon chemical potential,
since the hybrid star matter should be charge neutral,
the hadron and the quark surfaces degenerate to the charge neutral lines.
On the blue line, however, the hadron and the quark matter can be both charged as long as their charges could be canceled.

Fig.~\ref{fig:HadronQuarkSurface} shows distinctly that, with the increasing of baryon chemical potential,
the pressure of the matter changes along the red dotted line,
until the red dotted line intersects with the blue line,
and the phase transition begins.
The pressure then changes along the blue line.
As the blue line intersects with the orange line,
the phase transition ends,
and the pressure changes along the orange line of quark matter.

\begin{figure}[!htb]
\vspace*{-2.5mm}
\includegraphics[width=0.40\textwidth]{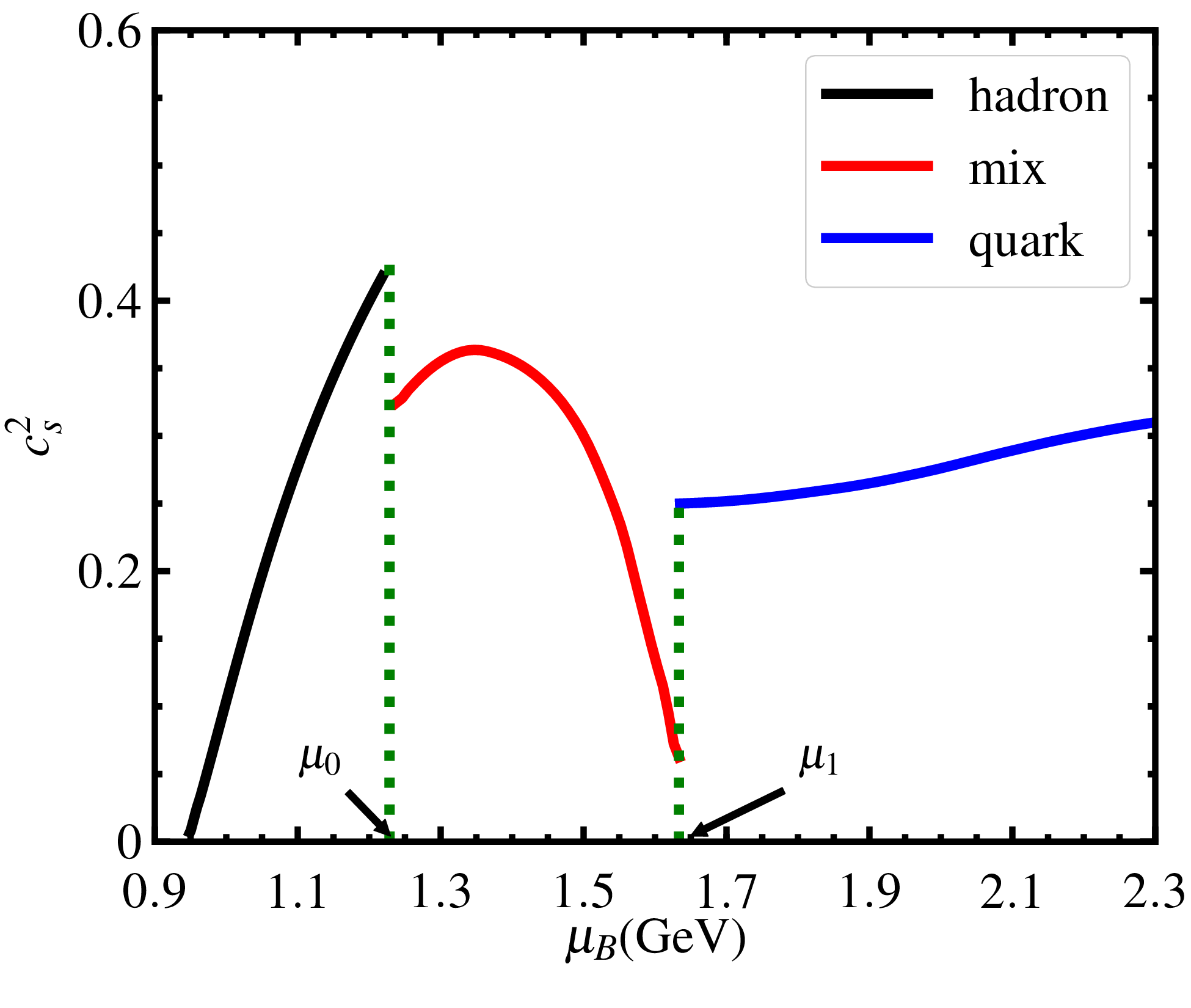}
\vspace*{-3.5mm}
\caption{(color online) General behavior of the sound speed squared as a function of baryon chemical potential.
The hadron matter (composing of protons, neutrons and leptons) is described with the TW-99 model in the RMF theory,
the quark matter is described with the DS equation of QCD with $\alpha=2$.
The $c_{s}^{2}$ of the mixed phase is obtained with Eq.(\ref{eqn:SoundSpeedDefinition}) whose input is the EoS gained with the Gibbs construction.}
\label{fig:mu-c2_Gibbs}
\vspace*{-2mm}
\end{figure}

On these lines, both the $p$ and the $\mu_{e}^{}$ can be represented by $\mu_{B}^{}$,
and the first order derivative of the pressure is:
\begin{equation}
 \frac{\textrm{d}p}{\textrm{d}\mu_{B}^{}}=\frac{\partial p}{\partial\mu_{B}^{}}+\frac{\partial p}{\partial\mu_{e}^{}}\frac{\textrm{d}\mu_{e}^{}}{\textrm{d}\mu_{B}^{}}.
\end{equation}
At the intersection point of the red dotted line and the blue line (denoted by blue balls in Fig.~\ref{fig:HadronQuarkSurface}),
the derivative ${\textrm{d}\mu_{e}^{}}/{\textrm{d}\mu_{B}^{}}$ is not continuous since the connection of the two lines is not smooth.
However, since $\partial p/{\partial \mu_{e}^{}}$ is zero on the red line because of the charge neutral requirement,
the first order derivative of the pressure is continuous on this cross point.
%
%Notice that the first derivative of pressure is discontinuous at the phase transition point under the Maxwell %construction where there is only one chemical potential.
%
The second order derivative ${\textrm{d}^{2}p}/{\textrm{d}\mu_{B}^{2}}$, however, is not continuous on the phase transition point,
and the square of the sound speed
\begin{equation}\label{eqn:SoundSpeedDefinition}
c_{s}^{2} = \frac{\textrm{d} p}{\textrm{d}\varepsilon} = \frac{\textrm{d}p/\textrm{d}\mu_{B}}{\mu_{B}\textrm{d}^{2} p/\textrm{d}\mu_{B}^{2}} \, ,
\end{equation}
is not continuous at this point.
The same goes for the intersection point of the blue and orange lines.
The general characteristic of the sound speed squared as a function of baryon chemical potential can be shown in Fig.~\ref{fig:mu-c2_Gibbs}.

It is known that, when the baryon chemical potential is barely beyond the proton mass,
the matter is in hadron phase, and the SS increases monotonically with the increasing of baryon chemical potential.
At some critical chemical potential $\mu_{0}$ (e.g., here $\mu_{0}^{}=1.23\,\textrm{GeV}$),
the quark matter begins to appear, and the sound speed squared gets discontinuous.
At another critical chemical potential $\mu_{1}$ (e.g., here $\mu_{1}=1.63\,\textrm{ GeV}$),
the SS involves another gap
where hadron matter disappears completely and the hadron-quark phase transition ends.
And in case of the chemical potential $\mu > \mu_{1}$, the $c_{s}^{2}$ increases monotonously and approaches the conformal limit $1/3$.
As for the $c_{s}^{2}(\mu_{B}^{})$ in the region $\mu_{0}^{} < \mu_{B}^{} < \mu_{1}^{}$, one still does not have any concrete knowledge.
It is just what we take as the objective to construct the EoS of the matter involving the hadron-quark phase transition.

\subsection{Maxwell, Gibbs and 3-window Constructions}\label{sec:OldConstruction}

In order for the hadron-quark phase transition described above to happen,
the hadron surface and the quark surface must have a cross line,
and this intersection line must also intersect with the charge neutral lines on both surfaces.

\begin{figure}[!htb]
\vspace*{-2.5mm}
\includegraphics[width=0.40\textwidth]{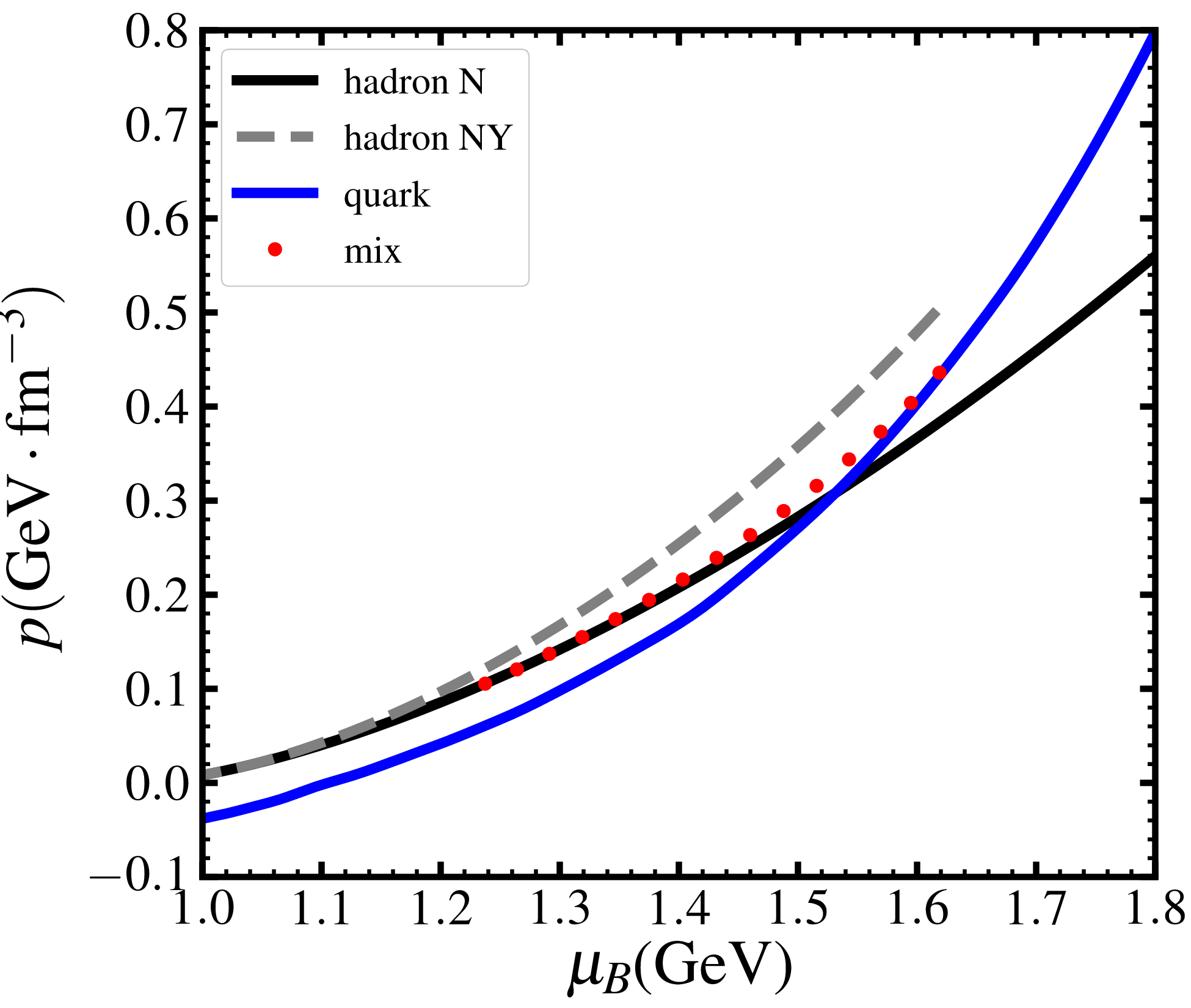}
\vspace*{-3.5mm}
\caption{(color online) Calculated pressure as a function of baryon chemical potential for different phases.
The black solid line corresponds to the hadron matter where only proton and neutron (and lepton) are included,
the gray line corresponds to the hadron matter which includes also hyperons,
and the blue solid line corresponds to the quark matter.
The red dotted line is the projection of the cross line in Fig.~\ref{fig:HadronQuarkSurface}.
The model we use are all described in the appendix.
}\label{fig:p_muB}
\vspace*{-2mm}
\end{figure}

In practice, it is not necessary (and in many cases, not possible)
to calculate the whole $p\,$--$\mu_{B}^{}$--$\mu_{e}^{}$ surfaces.
What we need to calculate are the charge neutral lines for both hadron and quark phases,
and project the obtained results onto the $p\,$--$\mu_{B}^{}$ plane.
In Fig.~\ref{fig:p_muB}, we present an example for such a projection.

It has been well known that, at low density, hadron phase is the stable phase.
The pressure of the hadron phase should then be larger than that of the quark phase.
At high density, the quark phase should have larger pressure to be more stable.
Therefore, for the phase transition to happen,
there must be a cross point for the $p\,$--$\mu_{B}^{}$ curves of the two phases.
The cross point, $(p_{0}^{},\mu_{B,0}^{})$, is just the phase transition point in Maxwell construction.

Under the Maxwell construction, the matter with baryon chemical potential $\mu_{B} \le \mu_{B,0}$ can be consequently described by hadron model,
while the matter with $\mu_{B}>\mu_{B,0}$ can be described by quark model.
To manifest the phase transition, one usually take into account the following Eq.(\ref{eqn:maxwell_construction}).
\begin{equation}\label{eqn:maxwell_construction}
  \begin{split}
p=p^{H}\theta(\mu_{B,0}^{}-\mu_{B}^{})+p^{Q}\theta(\mu_{B}^{}-\mu_{B,0}^{}),\\
\varepsilon=\varepsilon^{H}\theta(\mu_{B,0}^{}-\mu_{B}^{})+\varepsilon^{Q}\theta(\mu_{B}^{}-\mu_{B,0}^{}),
  \end{split}
\end{equation}
where $p^{H}$, $\varepsilon^{H}$ is the pressure, the energy density for hadron matter,
and $p^{Q}$, $\varepsilon^{Q}$ is the pressure, the energy density for quark matter, respectively.
The $\theta(x)$ is the usual step function.

However, if we look at the $p\,$--$\mu_{B}$--$\mu_{e}$ space in Fig.~\ref{fig:HadronQuarkSurface} more carefully,
the Maxwell construction corresponds to a sudden jump from the charge neutral line of hadron phase to the charge neutral line of quark phase,
{\it i.e.}, the electron chemical potentials of the two phases are not the same at $\mu_{B,0}^{}$
and the phase transition condition (equal pressure and equal chemical potential) is actually not satisfied.

In order for the phase transition condition to be satisfied,
the phase should evolve along the cross line of the hadron and quark surfaces in Fig.~\ref{fig:HadronQuarkSurface},
which is projected as the red dotted line in Fig.~\ref{fig:p_muB}.
This red dotted line then describes the mixed phase where hadron matter and quark matter coexist.
We notice that this mixed phase is exactly the same as we get from usual Gibbs construction,
except that instead of the whole hadron and quark surfaces,
only the points on the cross line are included.
The usual equation for Gibbs construction reads:
\begin{equation}
  \begin{split}
    &p^{H}(\mu_{B},\mu_{e})=p^{Q}(\mu_{B},\mu_{e}),\\
    &(1-\chi) n_{c}^{H}(\mu_{B},\mu_{e}) = -\chi n^{Q}_{c}(\mu_{B},\mu_{e}),\\
  \end{split}
\end{equation}
where $n_{c}^{H}$, $n_{c}^{Q}$ is the charge density for hadron matter, quark matter, respectively.
$\chi$ is the quark volume fraction to be determined.
Then, the energy density of the mixed phase is:
\begin{equation}
  \varepsilon=(1-\chi)\varepsilon^{H}+\chi\varepsilon^{Q}.
\end{equation}

Both Maxwell and Gibbs constructions have been widely used
and succeeded in constructing hybrid stars which satisfy astronomical constraints (see, e.g., Refs.~\cite{Glendenning:2000,Oertel:2016bki,Wei:2018mxy,Husain:2020nbb,Miyatsu:2015kwa,Han:2019bub} and many others).
However, there are cases where they are not adaptable.
For example, in Fig.~\ref{fig:p_muB}, the gray dashed line corresponds to the hadron matter with the inclusion of hyperons,
and it does not intersect with the quark line in the cases $\mu_{B}^{} \lesssim 1.6\,$GeV.
A naive speculation by eye seems to suggest that there will be a cross point at larger chemical potential,
but the hadron model, such as the RMF, will break down at such a large density and there will be no physical solution to the equations.

This problem should not be regarded as a flaw of the construction, but instead is due to the inappropriate application.
One knows well that hadron models are calibrated by experiment at low density,
and are unreliable at high densities.
Similarly, the quark models should only be adopted in high density region,
and approach the asymptotic freedom with the increasing of the density,
they should then not be implemented at low densities.
At intermediate density where the phase transition takes place,
both hadron model and quark approach lose their accuracy.
Therefore, even if the $p(\mu_{B}^{})$ curve of the two phases do not intersect,
it does not mean that the phase transition will be prohibited.

In order to overcome the above-mentioned inaccuracy problem,
the 3-window construction was proposed~\cite{Masuda:2012ed,Masuda:2012kf,Baym:2017whm,Kojo:2014rca,Kojo:2015fua}.
In the spirit of 3-window construction,
one can get the EoS of the matter in the middle density region for the two phases to coexist by interpolating the results of the hadron and the quark models.
There have been several ways to do the interpolation.
For example, in Ref.~\cite{Masuda:2012ed}, the energy density is interpolated with weight function $f_{\pm}$:
\begin{equation}
  \begin{split}
  &\varepsilon=f_{-}\varepsilon^{H}+f_{+}\varepsilon^{Q},\\
  &f_{\pm}=\frac{1}{2}\left(1\pm\tanh\left(\frac{n_{B} - \overline{n}}{\Gamma}\right)\right),
  \end{split}
\end{equation}
where $n_{B}^{}$ is the baryon number density, $\overline{n}$ and $\Gamma$ is the parameter for the weight function.
The pressure and chemical potential can then be derived with general thermodynamical relations.

However, these interpolation schemes assume that the transition from hadron to quark matter is a crossover,
and have not yet taken into account the characteristics of the first--order phase transition which has been proposed to happen in cold dense strong interaction matter with continuum field theory of QCD (see, e.g., Refs.~\cite{Qin:2010nq,Bashir:2012fs,Gao:2016qkh,Fischer:2018sdj,Fu:2019hdw,Gao:2020qsj} and others).
Since the 3-window construction can be viewed as a generalization of Gibbs construction,
where the quark volume fraction is actually a special kind of weight function,
it is possible to find a way which combine the advantages of both the methods.

In this paper, by inspiration of the Maxwell, Gibbs and 3-window construction,
we propose a new construction scheme which describes a first-order phase transition with two conservation charges,
while avoid using the hadron and quark model in intermediate density region where they are not reliable enough.

\subsection{Sound Speed Construction}\label{sec:SoundSpeedConstruction}

In the spirit of 3-window construction, we propose a new scheme to interpolate the EoS in the phase transition region.
Using this method, the large uncertainties of hadron and quark theories will be absorbed into the uncertainties of the construction parameters.
We will see that the chemical potentials labelling the phase transition region can be taken as the construction parameters.
Then, it will be easier to constrain the phase transition chemical potentials directly from astronomical observations,
instead of constraining them indirectly through the constraints on hadron/quark theory parameters.

In order to address the first-order phase transition with two conservation charges,
we make use of the variation feature of the SS to construct the EoS.
It has been known that the speed of sound in hadron matter at low density and that in quark matter at high density can have been calculated with quite high precision via the respective approach, respectively.
In this work, we take the RMF model for hadron matter, and DS equation approach of QCD for quark matter.
Inspired by the result obtained via the Gibbs construction, illustrated in Fig.~\ref{fig:mu-c2_Gibbs},
we construct the square of SS in the middle density region for the hadron and quark phases to coexist with a polynomial function:
\begin{equation}\label{eqn:InterpFunction}
c_{M}^{2}(\mu_{B})= \sum_{n=0}^{N}a_{n} \mu_{B}^{\,n} \, ,
\end{equation}
where $\{a_{n}\}$ are parameters to be fixed.
The phase transition region is denoted as $\mu_{0}^{}\le\mu_{B}\le\mu_{1}^{}$,
where $\mu_{0}^{}$, $\mu_{1}^{}$ correspond to the beginning, the ending of the phase transition, respectively.

After constructing the SS as a function of baryon chemical potential
we can calculate the EoS of the matter by solving the ordinary differential equations (ODEs):
\begin{equation} \label{eqn:EoSandSS}
\frac{\partial n_{B}}{\partial {\mu_{B}^{}}} = \frac{n_{B}}{{\mu_{B}^{}}c_{s}^{2}({\mu_{B}^{}})},\qquad \quad
\frac{\partial p}{\partial {\mu_{B}^{}}} = n_{B} \, ,
\end{equation}
which are just simply the thermodynamic relations.
The initial condition is
\begin{equation}\label{eqn:boundary1}
 n_{B}^{}(\mu_{0}) = n_{H}^{}(\mu_{0}),\qquad \quad p(\mu_{0}) = p_{H}^{}(\mu_{0}) \, ,
\end{equation}
where $n_{H}$ and $p_{H}^{}$ is the baryon number density and the pressure calculated via the hadron model.

Similarly, concerning the phase equilibrium condition, we have another boundary condition:
\begin{equation}\label{eqn:boundary2}
p(\mu_{1})= p_{Q}^{}(\mu_{1})\,, \qquad \quad n_{B}^{} (\mu_{1}) = n_{Q}^{}(\mu_{1})\,,
\end{equation}
where $p_{Q}^{}$ and $n_{Q}^{}$ is the pressure, the baryon number density of the quark matter, respectively.
The construction requires $N+3$ parameters ($N+1$ for the polynomial function, and 2 for the boundary chemical potentials).
After taking Eq.~(\ref{eqn:boundary2}) to constrain the parameters,
only $N+1$ parameters are independent.

Instead of directly using $\{a_{n}\}$,
we choose a set of sound speed squared $c_{s,(i)}^{2}$, as free parameters.
Each sound speed squared corresponds to a certain chemical potential, $\mu_{B}^{(i)}$,
where $i=1,\cdots,N-1$.
We require that $\mu_{B}^{(i)}$ be evenly distributed in the range $(\mu_0,\mu_1)$.
For example, for $N=2$, $\mu_{B}^{(1)}=(\mu_{0} + \mu_{1})/2$,
and for $N=3$, we have $\mu_{B}^{(1)}=\mu_{0}+\frac{1}{3}(\mu_{1} - \mu_{0})$, and $\mu_{B}^{(2)}=\mu_{0}+\frac{2}{3}(\mu_{1} - \mu_{0})$.
Therefore, the free parameters are: $\mu_{0}$, $\mu_{1}$ and $c_{s,(1)}^{2}, \cdots ,c_{s,(N-1)}^{2}$.

In our calculation, the ranges of these parameters are set as:
$0.938\le \mu_{0} \le 1.5 \textrm{GeV}$,
$1.2 \textrm{GeV}\le \mu_{1} \le 2.0 \textrm{GeV}$, $0 \le c_{s,(i)}^{2} < 1.0$.
The range of the sound speed squared is set with the causality and stability (see, e.g., Refs.~\cite{Tews:2018kmu,Greif:2018njt,Han:2019bub}).
The lower bound of $\mu_{0}$ is proposed by considering naively the appearance of nuclear matter.
The other bounds of the chemical potential are preset arbitrarily,
but we will see later that these ranges are irrelevant since they will be further narrowed down by astronomical observations.
We randomly choose these parameters from their corresponding ranges
and construct the sound speed squared,
then integrate Eq.(\ref{eqn:EoSandSS}) to get the EoS in each case.
We do this for 200000 times, and get the values for the SS and EoS for each set of parameters.

Before applying astronomical restrictions, some parameter sets should be automatically abandoned.
The parameter sets with $\mu_{0}>\mu_{1}$ will not be used, since it is physically meaningless.
Also, it is necessary to require that at chemical potential $\mu_{0}$ and $\mu_{1}$,
the SS of mixed phase is smaller than that of the hadron phase, the quark phase, respectively,
as shown in Fig.~\ref{fig:mu-c2_Gibbs}.
The parameter sets which do not satisfy this requirement will be abandoned.
The reason can be seen from Fig.~\ref{fig:p_muB}.
In order for the mixed phase to be favored, it must have a larger pressure than both hadron and quark phase in the range $\mu_{0}<\mu_{B}<\mu_{1}$.
Since the pressure $p$ and its first order derivative $\text{d}p/\text{d}\mu_{B}$ is continuous at the phase transition chemical potentials,
the mixed phase must have a larger second order derivative to have a larger pressure.
Then, from Eq.(\ref{eqn:SoundSpeedDefinition}), we can see that this larger second order derivative will lead to a smaller sound speed squared.

\section{Numerical Calculation, Results and Discussions}\label{sec:numerical}

After having the energy density $\varepsilon$ and pressure $p$ of the cold dense matter,
we can calculate the mass-radius relation of the compact star
by solving the TOV equation (see, e.g., Ref.~\cite{Glendenning:2000})
\begin{eqnarray}
		\frac{\textrm{d}p}{\textrm{d}r} & =  & -\frac{m\varepsilon}{r^{2}}\frac{(1+p/\varepsilon)(1+4\pi r^{3}p/m)}{1-m/r}, \nonumber \\
		\frac{\textrm{d}m}{\textrm{d}r} & =  &  4\pi r^{2}\varepsilon, \nonumber
\end{eqnarray}
where we have taken the $G=1$ unit.

The tidal deformability $\Lambda$ of the star can be related to the star's mass $M$ and radius $R$ as
%\begin{equation}
$$ \Lambda = \frac{2}{\,3\,}k_{2}\left(\frac{R}{M}\right)^{5}\,, $$
%\end{equation}
%
where $k_{2}$ is the tidal love number of the compact star, which can be calculated
as~\cite{Flanagan:2007ix,Zhao:2018nyf} % ,Hinderer:2008ApJ,Postnikov:2010PRD}
\begin{eqnarray}
k_{2} & = & \frac{8}{5}\beta^{5}z \big{[} 6\beta(2\! -\! y_{R}^{} )+6\beta^{2}(5 y_{R}^{} \!- \! 8) + 4\beta^{3}(13 \! - \! 11y_{R}^{} ) \nonumber \\
	  &   & +4\beta^{4}(3y_{R}^{}\! - \! 2) + 8\beta^{5}(1 \!+ \!y_{R}^{} ) + 3 z \textrm{ln}(1 \! -\! 2\beta) \big{]}^{-1} \, , \nonumber
\end{eqnarray}
where $\beta=M/R$ is the compactness parameter,
%\begin{equation}
$$	z\equiv(1-2\beta^{2})(2 - y_{R}^{} +2\beta(y_{R}^{} -1)) \, , $$
%\end{equation}
and $y_{R}^{} =y(r=R)$ where $y(r)$ is calculated by solving the equation
%\begin{equation}
$$	\frac{\textrm{d}y}{\textrm{d}r}=-\frac{y^{2}}{r}-\frac{y-6}{r-2m} - rQ \, , $$
%\end{equation}
with
\begin{equation}
\begin{split}
	Q =& \, 4\pi\frac{(5\! - \! y)\varepsilon+(9 \! + \! y) p + (\varepsilon \! + \! p)/c_{s}^{2}}{1\! - \! 2m/r}\\
		 & -\left[\frac{2(m \! + \! 4\pi r^{3}p)}{r(r-2m)}\right]^{2} .
\end{split}
\end{equation}

Since astronomical observations have provided the maximum mass, the tidal deformability and other information of the stars,
we then take these observations to constrain our constructed EoS.

Since neutron stars with masses over $2M_{\odot}$ has been observed (see, e.g., Refs.~\cite{Demorest:2010bx,Antoniadis:2013pzd} and others),
we should exclude the constructed EoSs which result in a maximum mass less than $2M_{\odot}$.
The detection of gravitational wave also sets the upper limit of the maximum mass of neutron stars,
so we require that the maximum mass is not larger than $2.16M_{\odot}$~\cite{Riley:2021pdl,Miller:2021qha,Margalit:2017dij,Rezzolla:2017aly,Ruiz:2017due}.

Based on the information provided by the gravitational wave detection for the tidal deformability $\Lambda$,
we require $\Lambda_{1.4}<800$ for a $1.4M_{\odot}$ neutron star according to Ref.~\cite{LIGOScientific:2017vwq},
and $\Lambda_{1.4}>120$ according to Ref.~\cite{Annala:2017llu}.

The radius of the neutron star is closely related to their tidal deformability.
However, the possible radius can still vary in a large range for some given $\Lambda$~\cite{Raithel:2022efm}.
Therefore, we also take the radius to constrain our construction.
Since we still do not have the model independent result for the neutron star radius, and neither the observations,
we implement $R \in (9.9,13.6)\,\textrm{km}$ for $1.4M_{\odot}$ neutron star according to Ref.~\cite{Annala:2017llu} to constrain the EoS.

After constraining the mass, radius and tidal deformability,
the baryon chemical potential region for the hadron-quark phase transition to occur can be constrained by the astronomical observations.

\subsection{For Quadratic Construction}\label{sec:Quadratic}

We first take a quadratic function to construct the sound speed squared,
that is, taking $N=2$ in Eq.~(\ref{eqn:InterpFunction}).

\begin{figure}[!htb]
\includegraphics[width=0.43\textwidth]{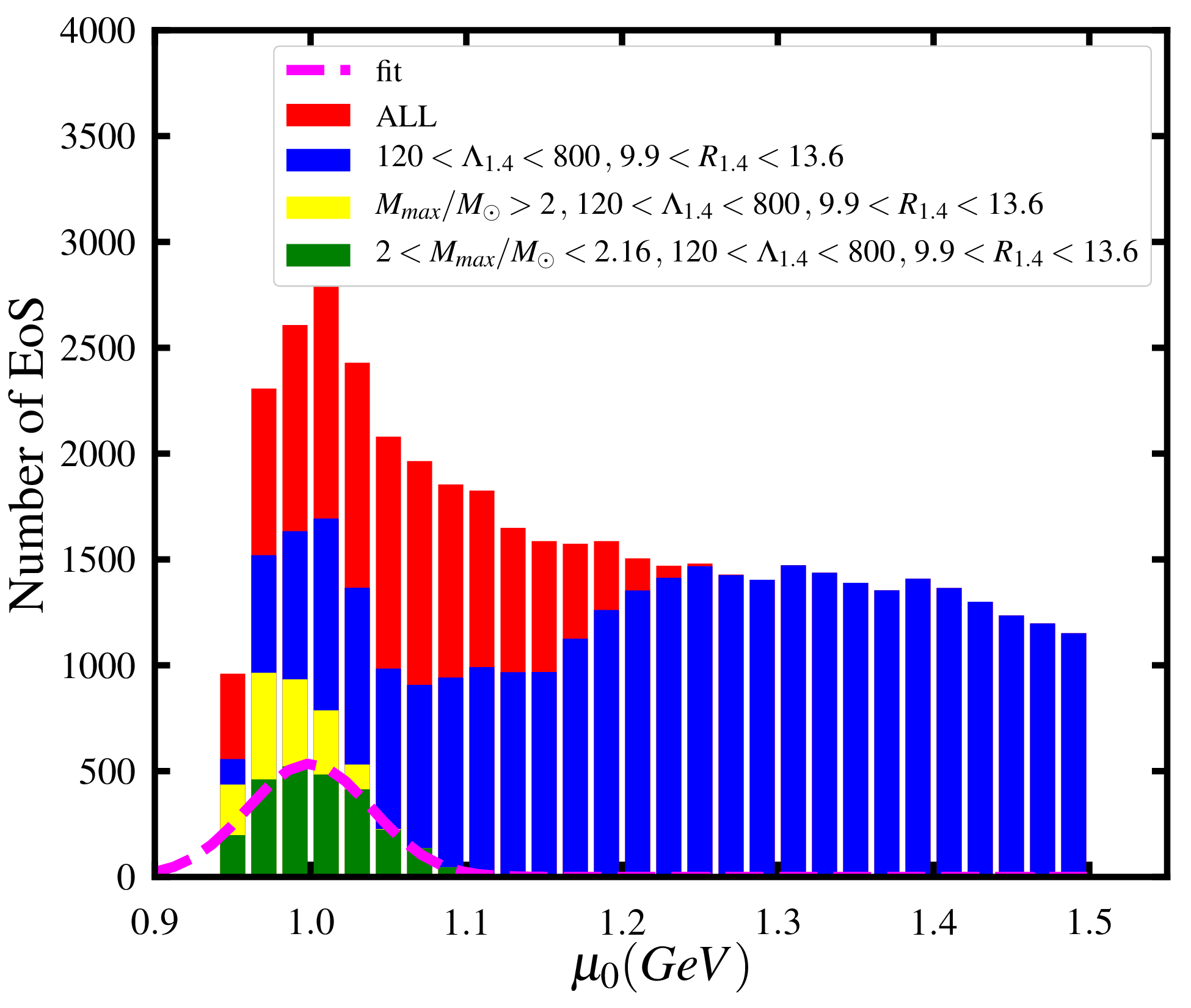}
\vspace*{-3mm}
\caption{(color online) Calculated $\mu_{0}^{}$ dependence of the number of constructed EoSs with several kinds of astronomical constraints.
In the construction, the hadron matter is described with the TW-99 model in the framework of the RMF with nucleons and hyperons,
and the quark matter is described via the DS equation approach of QCD with $\alpha=2$.
}
\label{fig:mu0_freq_NYq}
\end{figure}

In Fig.~\ref{fig:mu0_freq_NYq}, we show the $\mu_{0}$ dependence of the number of the EoSs satisfying several kinds of astronomical constraints.
The hadron matter is described with the RMF model with the inclusion of hyperons,
the quark matter is described with the DS equation approach of QCD,
and the mixed phase region is constructed using the SS interpolation introduced in Sec.~\ref{sec:SoundSpeedConstruction} with a quadratic function.
We show the number of the EoSs without any astronomical constraints with red bars,
those satisfying the requirement $120<\Lambda_{1.4}<800$, $R_{1.4} \in (9.9,13.6)\,\textrm{km}$ with blue bars,
those with further constraint $M_{\textrm{max}} > 2 M_{\odot}$ with yellow bars,
and the ones with much further requirement $ 2 M_{\odot} < M_{\textrm{max}} < 2.16M_{\odot}$ in green bars.
Although the parameters are taken randomly in the corresponding ranges,
the red bars which label the number of the EoSs without any astronomical constraint shown in Fig.~\ref{fig:mu0_freq_NYq} is not uniformly distributed.
This is because when $\mu_{0}$ is too large,
there are possibilities for $\mu_{0} > \mu_{1}$,
or the construction cannot satisfy the boundary condition in Eq.~(\ref{eqn:boundary2}).

From Fig.~\ref{fig:mu0_freq_NYq}, one can find easily that, although the constraints on the tidal deformability
and the radius reduce the number of the EoSs for different values of the $\mu_{0}$,
they do not change the range of the $\mu_{0}$.
Meanwhile, the lower limit of the maximum mass reduces the upper limit of the $\mu_{0}$,
but the upper limit of the maximum mass doesn't change the range of $\mu_{0}$.

After taking into account all the astronomical constraints,
the range of $\mu_{0}$ is constrained to be $\mu_{0}\le 1.12\,\textrm{GeV}$,
which corresponds to a baryon number density $n_{0}\le 3.16\,n_{s}$,
where $n_{s}=0.153\,\textrm{fm}^{-3}$ is the saturation nuclear matter density.
However, the lower limit of $\mu_{0}$ is only $0.938\,\textrm{GeV}$,
corresponding to nearly zero baryon number density
which is consistent with the DS equation result given in Ref.~\cite{Gao:2016qkh}.

Since our initial parameters are randomly distributed in their corresponding ranges,
the number of the EoSs should be proportional to the probability distribution.
The distribution of the green bar is very similar to the normal distribution,
except that there is no data points for $\mu_{0} <0.938\,$GeV.

The normal distribution has the form:
\begin{equation}\label{eqn:NormalDistribution}
\begin{split}
&p(\mu\le x\le \mu+\text{d}\mu)\\
\equiv&\,f(\mu;E,\sigma)\text{d}\mu
=\frac{1}{\sqrt{2\pi \sigma^2}}\exp\left[-\frac{(\mu-E)^2}{2\sigma^2}\right]\textrm{d}\mu,
\end{split}
\end{equation}
where $p(\mu\le x\le \mu+\text{d}\mu)$ is the probability for a data to be in the range $(\mu,\mu+\text{d}\mu)$,
$E$ and $\sigma$ is the expectation value and the standard deviation, respectively.

By integrating the normal distribution function, we get the probability for a data to be in the range $(-\infty,\mu)$:
\begin{equation}
	\begin{split}
P(x\le\mu) &= \int_{-\infty}^{\mu}f(x;E,\sigma) \textrm{d} x =\frac{1}{2}\left[1+\text{erf}\left(\frac{\mu-E}{\sqrt{2}\sigma}\right)\right]\\
&\equiv F(\mu;E,\sigma).
	\end{split}
\end{equation}

Since we need to neglect all the data with $x\le \mu_c=0.938\,$GeV,
we should then reformulate the function as:
\begin{equation}
P(\mu_{c} \le x\le \mu) %\\
=\frac{F(\mu) - F(\mu_{c})}{1 - F(\mu_{c})} \equiv \overline{F}(\mu) \, .
\end{equation}

Using function $\overline{F}$ to fit the data-set of $\mu_{0}$,
we can get $E$ and $\sigma$ with the least square fitting.
The distribution function then follows $f(\mu_{0};E,\sigma)$ which is defined in Eq.~(\ref{eqn:NormalDistribution})
(except for the cut-off).

The fitted $f(\mu_{0};E,\sigma)$ is displayed in Fig.~\ref{fig:mu0_freq_NYq} with pink dashed line.
The most probable value of the $\mu_{0}$ is $\langle\mu_{0}\rangle = E =1.00\,\textrm{GeV}$,
with the standard fitting error being $5.3\times 10^{-5}\,$GeV.
This most probable chemical potential corresponds to a baryon number density $1.64\,n_{s}$
at which the nucleons in the matter begin to overlap with each other~\cite{Liu:2001em}.

\begin{figure}[!htb]
\includegraphics[width=0.42\textwidth]{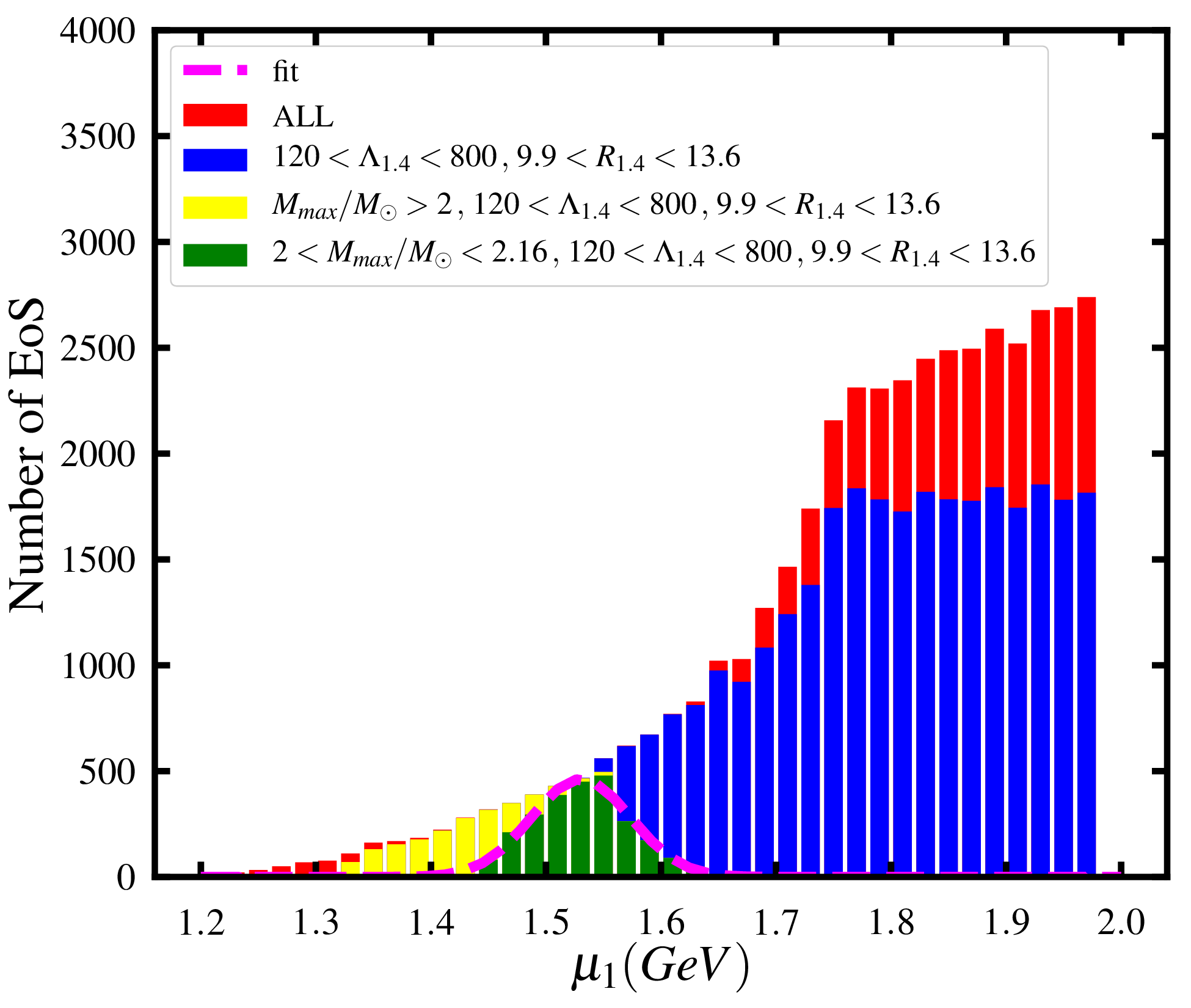}
\vspace*{-3mm}
\caption{(color online) The same as Fig.~\ref{fig:mu0_freq_NYq} except that this is for the $\mu_{1}$ dependence.
}
\label{fig:mu1_freq_NYq}
\end{figure}

Similar analysis has been carried out on $\mu_{1}$,
the baryon chemical potential corresponding to the ending of the hadron--quark phase transition.
The obtained $\mu_{1}$ dependence of the number of EoSs is shown in Fig.~\ref{fig:mu1_freq_NYq}.
As can be seen from the figure,
after applying the astronomical constraints,
the range of the $\mu_{1}$ is assigned as $1.42\le \mu_{1}\le 1.65\,\textrm{GeV}$,
corresponding to a baryon number density range $6.13\le n_{1}/n_{s}\le 11.14$.
As the same as done for $\mu_{0}$, we fit the distribution of EoSs with a normal distribution function,
and find that the most probable chemical potential is $\langle\mu_{1}\rangle=1.53\,\textrm{GeV}$,
where the standard fitting error is $2.7\times 10^{-5}\,$GeV.
This chemical potential corresponds to baryon density $\langle n_{1}\rangle =8.22\,n_{s}$.

%\begin{onecolumn}
\begin{table*}[htb]
\begin{center}
\caption{Constrained quantities featuring the hadron-quark phase transition under different astronomical observations.
The composition Nq, NYq refers to the case that the hadron matter sector does not include,
or includes hyperons, respectively.
The baryon chemical potentials are in the unit of GeV, and the baryon number densities are in the unit of $n_{s}$.}
\label{tab:other_constraints}
\vspace*{1mm}
\begin{tabular}{|c|c|c|c|c|c|c|c|c|c|c|c|c|c|}
\hline
\multicolumn{3}{|c|}{Astronomical observations} & &\multicolumn{10}{c|}{Constrained Range of the quantities}\\
\hline
$M_{\textrm{max}}(M_{\odot})$ & $\Lambda_{1.4}$&$R_{1.4}(\textrm{km})$ & composition & $\mu_{0\textrm{max}}$ & $n_{0\textrm{max}}$ & $\langle\mu_{0}\rangle$ & $\langle n_{0}\rangle$&$\mu_{1\textrm{min}}$&$\mu_{1\textrm{max}}$&$n_{1\textrm{min}}$&$n_{1\textrm{max}}$&$\langle\mu_{1}\rangle$&$\langle n_{1}\rangle$\\
\hline
	\multirow{2}{*}{2-2.16}& \multirow{2}{*}{120-800}&\multirow{2}{*}{9.9-13.6}&Nq &1.32&4.11&1.00 &1.57 &1.42&1.66&6.06&11.37&1.52&8.07 \\
						   &                         &                         &NYq&1.12&3.16&1.00 &1.57 &1.42&1.65&6.13&11.14&1.53&8.22 \\
\hline
	\multirow{2}{*}{2-2.35}& \multirow{2}{*}{120-800}&\multirow{2}{*}{9.9-13.6}&Nq &1.32&4.11&1.00 &1.57 &1.31&1.66&4.57&11.36&1.50&7.64  \\
						   &                         &                         &NYq&1.12&3.16&0.99 &1.38 &1.31&1.65&4.56&11.14&1.50&7.59  \\
\hline
	\multirow{2}{*}{2-2.16}& \multirow{2}{*}{344~\cite{Malik:2018zcf}-800}&\multirow{2}{*}{9.9-13.6}&Nq &1.32&4.11&1.08 &2.39 &1.42&1.58&6.06&9.25 &1.50&7.62  \\
						   &                         &                                              &NYq&1.05&2.31&0.99 &1.35 &1.42&1.59&6.13&9.52 &1.49&7.39   \\
\hline
\multirow{2}{*}{2-2.16}& \multirow{2}{*}{120-800}&\multirow{2}{*}{10.62-12.83~\cite{Lattimer:2014sga}}&Nq &1.32&4.11&1.04&1.99&1.42&1.61&6.06&10.13&1.51&7.84   \\
						   &                         &                                                 &NYq&1.12&3.16&1.01&1.63&1.42&1.63&6.13&10.72&1.52&8.07     \\
%\hline
%	\multirow{2}{*}{2-2.16}& \multirow{2}{*}{344-580~\cite{LIGOScientific:2018cki}}&\multirow{2}{*}{10.62-12.83} & Nq&1.32&4.11&1.08&2.39&1.42&1.58&6.06&9.24 &1.50&7.62\\
%						   &                         &                                                   &NYq&1.05&2.31&0.99&1.39&1.44&1.59&6.40&9.52 &1.49&7.43\\
\hline
	\multirow{2}{*}{2-2.16}& \multirow{2}{*}{344-580~\cite{LIGOScientific:2018cki}}&\multirow{2}{*}{10.4-11.9~\cite{Capano:2019eae}} &Nq&1.34&4.26&1.09&2.51&1.44&1.58&6.54&9.35 &1.50&7.70\\
						   &                         &                                                   & NYq&1.05&2.30&0.99&1.36&1.45&1.58&6.56&9.24 &1.50&7.53\\
\hline
\end{tabular}
\end{center}
\vspace*{-5mm}
\end{table*}
%
%\end{onecolumn}

We have shown above that astronomical observations can constrain the chemical potential region of the phase transition.
However, the values of the astronomical observables, e.g., the mass, the tidal deformability and the radius
have not yet been fixed concretely in observations, and different studies give distinct results.
Therefore, we repeat the above described analyzing process with different astronomical constraints,
and for each set of the constraints,
we consider two different cases: hyperons appear or do not appear in the hadron matter.
The obtained main characteristics of the results are listed in Table~\ref{tab:other_constraints}.

The first set of Table~\ref{tab:other_constraints} gives the result we have just described above.
The notation Nq, NYq refers to the case that the hadron matter sector does not include, or includes hyperons, respectively.
Comparing the results with and without hyperons,
one can observe that the upper limit of the $\mu_{0}$ decreases after including hyperons.
The second set lists the result when a larger upper limit of the maximum mass of a star is taken.
It is evident that, the lower limit of the $\mu_{1}$ reduces a lot, no matter whether hyperons are included.
The third set shows the result under a larger lower limit of the tidal deformability.
It manifests clearly that the upper limits of the $\mu_{1}$ is reduced.
Meanwhile, the upper limit of the $\mu_{0}$ is also reduced when hyperons are included.
The fourth set shows the result under a smaller radius range.
One can notice from the data that the upper limit of the $\mu_{1}$ is reduced,
but the change is smaller comparing to the second and the third set.
This means that the constraints from the mass and the tidal deformability is more stringent.

The obtained result under the most strict astronomical constraints is given in the last set of Table~\ref{tab:other_constraints}.
It is apparent that the regions of the $\mu_{0}$ and the $\mu_{1}$ are narrowed down correspondingly.
Furthermore, under these constraints,
the most probable beginning and ending chemical potentials are $\langle\mu_{0}\rangle=1.09\,\textrm{GeV}$, $\langle\mu_{1}\rangle=1.50\,\textrm{GeV}$ in the case without hyperons,
and $\langle\mu_{0}\rangle=0.99\,\textrm{GeV}$ and $\langle\mu_{1}\rangle=1.50\,\textrm{GeV}$
when hyperons are included.
It indicates that the chemical potential corresponding to the ending of the hadron--quark phase transition is not influenced much by the inclusion of hyperons under such an astronomical circumstance,
but the chemical potential corresponding to the beginning of the phase transition is reduced obviously by hyperons.

\subsection{For Cubic Construction}\label{sec:Cubic}

In previous subsection, we have shown the results with the EoS which generates the SS with a quadratic function in the mixed phase region,
i.e., taking $N=2$ in Eq.(\ref{eqn:InterpFunction}),
and find the most probable value for the phase transition to take place.
To further verify our result, we will consider the cubic interpolation function for the SS in the mixed phase region,
i.e., taking $N=3$ in Eq.(\ref{eqn:InterpFunction}) in this subsection.

Taking the same technique described in Sec.~\ref{sec:Quadratic},
we can find the most probable values of the baryon chemical potential region for the hadron-quark phase transition to happen under the cubic SS interpolation.
The obtained results and the comparison with those via the quadratic interpolation are shown in Table~\ref{tab:QuadraticCubicComparison}.

\begin{table}[!htbp]
\begin{center}
\caption{A comparison between the obtained results of the $\langle \mu_{0}\rangle$, $\langle \mu_{1} \rangle$ and their errors (in unit GeV) via the quadratic and cubic interpolation functions.
The astronomical constraints implemented are: $2.0\le M\le 2.16\,M_{\odot}$, $120\le\Lambda \le 800$ and $9.9\le R\le 13.6\,$km.
The ``err" columns list the standard error for the fitting of truncated normal distribution.}
\label{tab:QuadraticCubicComparison}
\vspace*{1mm}
\begin{tabular}{|c|c|c|c|c|c|}
\hline
Composition     &Construct& $\langle\mu_{0}\rangle$ & err. $\mu_{0}$    & $\langle\mu_{1}\rangle$ & err. $\mu_{1}$\\
\hline
\multirow{2}{*}{N} &quadratic&          1.00           &$3.7\!\times \!10^{-4}$&        1.52             &$2.7\! \times \! 10^{-5}$\\
                   &cubic    &          0.99           &$4.1\!\times \!10^{-4}$&        1.54             &$2.2\! \times \! 10^{-5}$\\
\hline
\multirow{2}{*}{NY}&quadratic&          1.00           &$5.4\! \times \! 10^{-5}$&        1.53             &$2.7\! \times \! 10^{-5}$\\
                   &cubic    &          1.00           &$5.4\! \times \! 10^{-5}$&        1.54             &$2.0\! \times \! 10^{-5}$\\
\hline
\end{tabular}
\end{center}
\vspace*{-5mm}
\end{table}

From the Table~\ref{tab:QuadraticCubicComparison}, one can observe evidently that
the most probable values labelling the baryon chemical potential region for the phase transition to take place is only very slightly changed by increasing the power of the interpolating function.
This fact indicates that with our construction scheme,
the region of the baryon chemical potential for the hadron-quark phase transition to take place (\textit{i.e.}, for the two phases to coexist) can be totally determined by astronomical observations,
regardless of the detail of the construction.

\section{\label{sec:sum}Summary and Remarks}

In this paper, we propose a scheme to interpolate the hadron and the quark models to construct
a complete EoS for the compact star matter involving hadron--quark phase transition.
We show the characteristic of the SS in the matter with two conservation charges and a first--order phase transition,
and take it as the objective to be interpolated.
To describe the hadron matter we take the RMF model with the TW-99 parameters,
and for the quark matter we implement the DS equation approach of QCD.

With the astronomical observations such as the maximum mass of the neutron star,
the radius and the tidal deformability of the star with  $M = 1.4\,M_{\odot}$ being taken as calibrations,
the baryon chemical potentials which correspond to the beginning and the ending of the hadron--quark phase transition are constrained to a quite small range.
Meanwhile, the distribution of the phase transition chemical potentials can be fitted with the normal distribution,
and the most probable values of the phase transition chemical potential and baryon number density are found.
The obtained results agree with the ones given via effective field theory of QCD very well.

We have also looked over the effect of the maximum mass, the radius, and the tidal deformability
on the phase transition chemical potentials by varying the calibration ranges of the observables.
It shows that a narrower range of the astronomical values indeed leads to a narrower range of the phase transition chemical potential.
With the most strict observation constraints, the most probable values of the hadron--quark phase transition
chemical potential and the baryon density are: $\langle\mu_{0}\rangle=0.99\,\textrm{GeV}$ and $\langle\mu_{1}\rangle=1.50\,\textrm{GeV}$, i.e., $\langle n_{0}\rangle=1.36\,n_{s}$
and $\langle n_{1}\rangle=7.53\, n_{s}$.

In order to verify the interpolation scheme,
we compared the results obtained with the quadratic and the cubic interpolation functions.
The results agree with each other excellently, indicating that our construction scheme is reliable.

Even though the phase transition chemical potentials have not yet been constrained to concrete values exactly now due to the astronomical observations having not provided exact values for the calibration quantities,
we have shown that the presently proposed scheme to determine the baryon chemical potential region for the hadron--quark phase transition to occur is efficient and found the most probable values.
With future detections, the maximum mass, the tidal deformability and the radius of neutron stars
can be measured with higher accuracy,
the range of the phase transition chemical potentials can be determined more precisely.

\begin{acknowledgments}
The work was supported by the National Natural Science Foundation of China under Contract Nos.\ 11435001, 11775041, 12175007,
12247107, and 12205353,
China Postdoctoral Science Foundation under Grant No. 2022M723230,
and CAS Project for Young Scientists in Basic Research (YSBR060).
\end{acknowledgments}

%\bibliography{reflib}
%\bibliographystyle{plain}
%\bibliography{Ref-SoundSpeedInterp}

\begin{thebibliography}{87}%
\makeatletter
\providecommand \@ifxundefined [1]{%
 \@ifx{#1\undefined}
}%
\providecommand \@ifnum [1]{%
 \ifnum #1\expandafter \@firstoftwo
 \else \expandafter \@secondoftwo
 \fi
}%
\providecommand \@ifx [1]{%
 \ifx #1\expandafter \@firstoftwo
 \else \expandafter \@secondoftwo
 \fi
}%
\providecommand \natexlab [1]{#1}%
\providecommand \enquote  [1]{``#1''}%
\providecommand \bibnamefont  [1]{#1}%
\providecommand \bibfnamefont [1]{#1}%
\providecommand \citenamefont [1]{#1}%
\providecommand \href@noop [0]{\@secondoftwo}%
\providecommand \href [0]{\begingroup \@sanitize@url \@href}%
\providecommand \@href[1]{\@@startlink{#1}\@@href}%
\providecommand \@@href[1]{\endgroup#1\@@endlink}%
\providecommand \@sanitize@url [0]{\catcode `\\12\catcode `\$12\catcode
  `\&12\catcode `\#12\catcode `\^12\catcode `\_12\catcode `\%12\relax}%
\providecommand \@@startlink[1]{}%
\providecommand \@@endlink[0]{}%
\providecommand \url  [0]{\begingroup\@sanitize@url \@url }%
\providecommand \@url [1]{\endgroup\@href {#1}{\urlprefix }}%
\providecommand \urlprefix  [0]{URL }%
\providecommand \Eprint [0]{\href }%
\providecommand \doibase [0]{http://dx.doi.org/}%
\providecommand \selectlanguage [0]{\@gobble}%
\providecommand \bibinfo  [0]{\@secondoftwo}%
\providecommand \bibfield  [0]{\@secondoftwo}%
\providecommand \translation [1]{[#1]}%
\providecommand \BibitemOpen [0]{}%
\providecommand \bibitemStop [0]{}%
\providecommand \bibitemNoStop [0]{.\EOS\space}%
\providecommand \EOS [0]{\spacefactor3000\relax}%
\providecommand \BibitemShut  [1]{\csname bibitem#1\endcsname}%
\let\auto@bib@innerbib\@empty
%</preamble>
%
\bibitem [{\citenamefont {Ding}\ \emph {et~al.}(2015)\citenamefont {Ding},
  \citenamefont {Karsch},\ and\ \citenamefont {Mukherjee}}]{Ding:2015ona}%
  \BibitemOpen
  \bibfield  {author} {\bibinfo {author} {\bibfnamefont {H.~T.}\ \bibnamefont
  {Ding}}, \bibinfo {author} {\bibfnamefont {F.}~\bibnamefont {Karsch}}, \ and\
  \bibinfo {author} {\bibfnamefont {S.}~\bibnamefont {Mukherjee}},\ }
  \href {\doibase 10.1142/S0218301315300076} {\bibfield  {journal} {\bibinfo
  {journal} {Int. J. Mod. Phys. E}\ }\textbf {\bibinfo {volume} {24}},\
  \bibinfo {pages} {1530007} (\bibinfo {year} {2015})},\ \Eprint
  {http://arxiv.org/abs/1504.05274} {arXiv:1504.05274} \BibitemShut {NoStop}%
%
\bibitem [{\citenamefont {Roberts}\ and\ \citenamefont
  {Schmidt}(2000)}]{Roberts:2000aa}%
  \BibitemOpen
  \bibfield  {author} {\bibinfo {author} {\bibfnamefont {C.~D.}\ \bibnamefont
  {Roberts}}\ and\ \bibinfo {author} {\bibfnamefont {S.~M.}\ \bibnamefont
  {Schmidt}},\ }\href {\doibase 10.1016/S0146-6410(00)90011-5} {\bibfield
  {journal} {\bibinfo  {journal} {Prog. Part. Nucl. Phys.}\ }\textbf {\bibinfo
  {volume} {45}},\ \bibinfo {pages} {S1} (\bibinfo {year} {2000})},\ \Eprint
  {http://arxiv.org/abs/nucl-th/0005064} {arXiv:nucl-th/0005064} \BibitemShut
  {NoStop}%
%
\bibitem [{\citenamefont {Qin}\ \emph {et~al.}(2011{\natexlab{a}})\citenamefont
  {Qin}, \citenamefont {Chang}, \citenamefont {Chen}, \citenamefont {Liu},\
  and\ \citenamefont {Roberts}}]{Qin:2010nq}%
  \BibitemOpen
  \bibfield  {author} {\bibinfo {author} {\bibfnamefont {S.~X.}\ \bibnamefont
  {Qin}}, \bibinfo {author} {\bibfnamefont {L.}~\bibnamefont {Chang}}, \bibinfo
  {author} {\bibfnamefont {H.}~\bibnamefont {Chen}}, \bibinfo {author}
  {\bibfnamefont {Y.~X.}\ \bibnamefont {Liu}}, \ and\ \bibinfo {author}
  {\bibfnamefont {C.~D.}\ \bibnamefont {Roberts}},\ }\href {\doibase
  10.1103/PhysRevLett.106.172301} {\bibfield  {journal} {\bibinfo  {journal}
  {Phys. Rev. Lett.}\ }\textbf {\bibinfo {volume} {106}},\ \bibinfo {pages}
  {172301} (\bibinfo {year} {2011}{\natexlab{a}})},\ \Eprint
  {http://arxiv.org/abs/1011.2876} {arXiv:1011.2876} \BibitemShut {NoStop}%
%
\bibitem [{\citenamefont {Bashir}\ \emph {et~al.}(2012)\citenamefont {Bashir},
  \citenamefont {Chang}, \citenamefont {Clo{\"e}t}, \citenamefont
  {{El-Bennich}}, \citenamefont {Liu}, \citenamefont {Roberts},\ and\
  \citenamefont {Tandy}}]{Bashir:2012fs}%
  \BibitemOpen
  \bibfield  {author} {\bibinfo {author} {\bibfnamefont {A.}~\bibnamefont
  {Bashir}}, \bibinfo {author} {\bibfnamefont {L.}~\bibnamefont {Chang}},
  \bibinfo {author} {\bibfnamefont {I.~C.}\ \bibnamefont {Clo{\"e}t}}, \bibinfo
  {author} {\bibfnamefont {B.}~\bibnamefont {{El-Bennich}}}, \bibinfo {author}
  {\bibfnamefont {Y.~X.}\ \bibnamefont {Liu}}, \bibinfo {author} {\bibfnamefont
  {C.~D.}\ \bibnamefont {Roberts}}, \ and\ \bibinfo {author} {\bibfnamefont
  {P.~C.}\ \bibnamefont {Tandy}},\ }\href {\doibase 10.1088/0253-6102/58/1/16}
  {\bibfield  {journal} {\bibinfo  {journal} {Commun. Theor. Phys.}\ }\textbf
  {\bibinfo {volume} {58}},\ \bibinfo {pages} {79} (\bibinfo {year} {2012})},\
  \Eprint {http://arxiv.org/abs/1201.3366} {arXiv:1201.3366} \BibitemShut
  {NoStop}%
%
\bibitem [{\citenamefont {Gao}\ and\ \citenamefont {Liu}(2016)}]{Gao:2016qkh}%
  \BibitemOpen
  \bibfield  {author} {\bibinfo {author} {\bibfnamefont {F.}~\bibnamefont
  {Gao}}\ and\ \bibinfo {author} {\bibfnamefont {Y.~X.}\ \bibnamefont {Liu}},\
  }\href {\doibase 10.1103/PhysRevD.94.076009} {\bibfield  {journal} {\bibinfo
  {journal} {Phys. Rev. D}\ }\textbf {\bibinfo {volume} {94}},\ \bibinfo {pages} {076009}  (\bibinfo {year}
  {2016})},\ \Eprint
  {http://arxiv.org/abs/1607.01675} {arXiv:1607.01675} \BibitemShut {NoStop}%
%
\bibitem [{\citenamefont {Fischer}(2019)}]{Fischer:2018sdj}%
  \BibitemOpen
  \bibfield  {author} {\bibinfo {author} {\bibfnamefont {C.~S.}\ \bibnamefont
  {Fischer}},\ }\href {\doibase 10.1016/j.ppnp.2019.01.002} {\bibfield
  {journal} {\bibinfo  {journal} {Prog. Part. Nucl. Phys.}\ }\textbf {\bibinfo
  {volume} {105}},\ \bibinfo {pages} {1} (\bibinfo {year} {2019})},\ \Eprint
  {http://arxiv.org/abs/1810.12938} {arXiv:1810.12938} \BibitemShut {NoStop}%
%
\bibitem [{\citenamefont {Herbst}\ \emph {et~al.}(2013)\citenamefont {Herbst},
  \citenamefont {Pawlowski},\ and\ \citenamefont {Schaefer}}]{Herbst:2013ail}%
  \BibitemOpen
  \bibfield  {author} {\bibinfo {author} {\bibfnamefont {T.~K.}\ \bibnamefont
  {Herbst}}, \bibinfo {author} {\bibfnamefont {J.~M.}\ \bibnamefont
  {Pawlowski}}, \ and\ \bibinfo {author} {\bibfnamefont {B.-J.}\ \bibnamefont
  {Schaefer}},\ }\href {\doibase 10.1103/PhysRevD.88.014007} {\bibfield
  {journal} {\bibinfo  {journal} {Phys. Rev. D}\ }\textbf {\bibinfo {volume}
  {88}},\ \bibinfo {pages} {014007} (\bibinfo {year} {2013})},\ \Eprint
  {http://arxiv.org/abs/1302.1426} {arXiv:1302.1426} \BibitemShut {NoStop}%
%
\bibitem [{\citenamefont {Fister}\ and\ \citenamefont
  {Pawlowski}(2013)}]{Fister:2013bh}%
  \BibitemOpen
  \bibfield  {author} {\bibinfo {author} {\bibfnamefont {L.}~\bibnamefont
  {Fister}}\ and\ \bibinfo {author} {\bibfnamefont {J.~M.}\ \bibnamefont
  {Pawlowski}},\ }\href {\doibase 10.1103/PhysRevD.88.045010} {\bibfield
  {journal} {\bibinfo  {journal} {Phys. Rev. D}\ }\textbf {\bibinfo {volume}
  {88}},\ \bibinfo {pages} {045010} (\bibinfo {year} {2013})},\ \Eprint
  {http://arxiv.org/abs/1301.4163} {arXiv:1301.4163} \BibitemShut {NoStop}%
%
\bibitem [{\citenamefont {Fu}\ \emph {et~al.}(2020)\citenamefont {Fu},
  \citenamefont {Pawlowski},\ and\ \citenamefont {Rennecke}}]{Fu:2019hdw}%
  \BibitemOpen
  \bibfield  {author} {\bibinfo {author} {\bibfnamefont {W.~J.}\ \bibnamefont
  {Fu}}, \bibinfo {author} {\bibfnamefont {J.~M.}\ \bibnamefont {Pawlowski}}, \
  and\ \bibinfo {author} {\bibfnamefont {F.}~\bibnamefont {Rennecke}},\ }\href
  {\doibase 10.1103/PhysRevD.101.054032} {\bibfield  {journal} {\bibinfo
  {journal} {Phys. Rev. D}\ }\textbf {\bibinfo {volume} {101}},\ \bibinfo
  {pages} {054032} (\bibinfo {year} {2020})},\ \Eprint
  {http://arxiv.org/abs/1909.02991} {arXiv:1909.02991} \BibitemShut {NoStop}%
%
\bibitem [{\citenamefont {Gao}\ and\ \citenamefont
  {Pawlowski}(2020)}]{Gao:2020qsj}%
  \BibitemOpen
  \bibfield  {author} {\bibinfo {author} {\bibfnamefont {F.}~\bibnamefont
  {Gao}}\ and\ \bibinfo {author} {\bibfnamefont {J.~M.}\ \bibnamefont
  {Pawlowski}},\ }\href {\doibase 10.1103/PhysRevD.102.034027} {\bibfield
  {journal} {\bibinfo  {journal} {Phys. Rev. D}\ }\textbf {\bibinfo {volume}
  {102}},\ \bibinfo {pages} {034027} (\bibinfo {year} {2020})},\ \Eprint
  {http://arxiv.org/abs/2002.07500} {arXiv:2002.07500} \BibitemShut
  {NoStop}%
%
\bibitem [{\citenamefont {Fukushima}\ and\ \citenamefont
  {Skokov}(2017)}]{Fukushima:2017csk}%
  \BibitemOpen
  \bibfield  {author} {\bibinfo {author} {\bibfnamefont {K.}~\bibnamefont
  {Fukushima}}\ and\ \bibinfo {author} {\bibfnamefont {V.}~\bibnamefont
  {Skokov}},\ }\href {\doibase 10.1016/j.ppnp.2017.05.002} {\bibfield
  {journal} {\bibinfo  {journal} {Prog. Part. Nucl. Phys.}\ }\textbf {\bibinfo
  {volume} {96}},\ \bibinfo {pages} {154} (\bibinfo {year} {2017})},\ \Eprint
  {http://arxiv.org/abs/1705.00718} {arXiv:1705.00718} \BibitemShut {NoStop}%
%
\bibitem [{\citenamefont {Fu}\ \emph {et~al.}(2008)\citenamefont {Fu},
  \citenamefont {Zhang},\ and\ \citenamefont {Liu}}]{Fu:2007xc}%
  \BibitemOpen
  \bibfield  {author} {\bibinfo {author} {\bibfnamefont {W.~J.}\ \bibnamefont
  {Fu}}, \bibinfo {author} {\bibfnamefont {Z.}~\bibnamefont {Zhang}}, \ and\
  \bibinfo {author} {\bibfnamefont {Y.~X.}\ \bibnamefont {Liu}},\ }\href
  {\doibase 10.1103/PhysRevD.77.014006} {\bibfield  {journal} {\bibinfo
  {journal} {Phys. Rev. D}\ }\textbf {\bibinfo {volume} {77}},\ \bibinfo
  {pages} {014006} (\bibinfo {year} {2008})},\ \Eprint
  {http://arxiv.org/abs/0711.0154} {arXiv:0711.0154} \BibitemShut {NoStop}%
%
\bibitem [{\citenamefont {Fukushima}(2008)}]{Fukushima:2008wg}%
  \BibitemOpen
  \bibfield  {author} {\bibinfo {author} {\bibfnamefont {K.}~\bibnamefont
  {Fukushima}},\ }\href {\doibase 10.1103/PhysRevD.77.114028} {\bibfield
  {journal} {\bibinfo  {journal} {Phys. Rev. D}\ }\textbf {\bibinfo {volume}
  {77}},\ \bibinfo {pages} {114028} (\bibinfo {year} {2008})},\ \Eprint
  {http://arxiv.org/abs/0803.3318} {arXiv:0803.3318} \BibitemShut {NoStop}%
%
\bibitem [{\citenamefont {Fukushima}(2012)}]{Fukushima:2011jc}%
  \BibitemOpen
  \bibfield  {author} {\bibinfo {author} {\bibfnamefont {K.}~\bibnamefont
  {Fukushima}},\ }\href {\doibase 10.1088/0954-3899/39/1/013101} {\bibfield
  {journal} {\bibinfo  {journal} {J. Phys. G}\ }\textbf {\bibinfo {volume}
  {39}},\ \bibinfo {pages} {013101} (\bibinfo {year} {2012})},\ \Eprint
  {http://arxiv.org/abs/1108.2939} {arXiv:1108.2939} \BibitemShut {NoStop}%
%
\bibitem [{\citenamefont {Gao}\ \emph {et~al.}(2016)\citenamefont {Gao},
  \citenamefont {Chen}, \citenamefont {Liu}, \citenamefont {Qin}, \citenamefont
  {Roberts},\ and\ \citenamefont {Schmidt}}]{Gao:2015kea}%
  \BibitemOpen
  \bibfield  {author} {\bibinfo {author} {\bibfnamefont {F.}~\bibnamefont
  {Gao}}, \bibinfo {author} {\bibfnamefont {J.}~\bibnamefont {Chen}}, \bibinfo
  {author} {\bibfnamefont {Y.~X.}\ \bibnamefont {Liu}}, \bibinfo {author}
  {\bibfnamefont {S.~X.}\ \bibnamefont {Qin}}, \bibinfo {author} {\bibfnamefont
  {C.~D.}\ \bibnamefont {Roberts}}, \ and\ \bibinfo {author} {\bibfnamefont
  {S.~M.}\ \bibnamefont {Schmidt}},\ }\href {\doibase
  10.1103/PhysRevD.93.094019} {\bibfield  {journal} {\bibinfo  {journal} {Phys.
  Rev. D}\ }\textbf {\bibinfo {volume} {93}},\ \bibinfo {pages} {094019}
  (\bibinfo {year} {2016})},\ \Eprint {http://arxiv.org/abs/1507.00875}
  {arXiv:1507.00875} \BibitemShut {NoStop}%
%
\bibitem [{\citenamefont {Brandes}\ \emph {et~al.}(2021)\citenamefont
  {Brandes}, \citenamefont {Kaiser},\ and\ \citenamefont
  {Weise}}]{Brandes:2021pti}%
  \BibitemOpen
  \bibfield  {author} {\bibinfo {author} {\bibfnamefont {L.}~\bibnamefont
  {Brandes}}, \bibinfo {author} {\bibfnamefont {N.}~\bibnamefont {Kaiser}}, \
  and\ \bibinfo {author} {\bibfnamefont {W.}~\bibnamefont {Weise}},\ }\href
  {\doibase 10.1140/epja/s10050-021-00528-2} {\bibfield  {journal} {\bibinfo
  {journal} {Eur. Phys. J. A}\ }\textbf {\bibinfo {volume} {57}},\ \bibinfo
  {pages} {243} (\bibinfo {year} {2021})},\ \Eprint
  {http://arxiv.org/abs/2103.06096} {arXiv:2103.06096} \BibitemShut {NoStop}%
%
\bibitem [{\citenamefont {Baym}\ \emph {et~al.}(2018)\citenamefont {Baym},
  \citenamefont {Hatsuda}, \citenamefont {Kojo}, \citenamefont {Powell},
  \citenamefont {Song},\ and\ \citenamefont {Takatsuka}}]{Baym:2017whm}%
  \BibitemOpen
  \bibfield  {author} {\bibinfo {author} {\bibfnamefont {G.}~\bibnamefont
  {Baym}}, \bibinfo {author} {\bibfnamefont {T.}~\bibnamefont {Hatsuda}},
  \bibinfo {author} {\bibfnamefont {T.}~\bibnamefont {Kojo}}, \bibinfo {author}
  {\bibfnamefont {P.~D.}\ \bibnamefont {Powell}}, \bibinfo {author}
  {\bibfnamefont {Y.~F.}\ \bibnamefont {Song}}, \ and\ \bibinfo {author}
  {\bibfnamefont {T.}~\bibnamefont {Takatsuka}},\ }\href {\doibase
  10.1088/1361-6633/aaae14} {\bibfield  {journal} {\bibinfo  {journal} {Rep.
  Prog. Phys.}\ }\textbf {\bibinfo {volume} {81}},\ \bibinfo {pages} {056902}
  (\bibinfo {year} {2018})},\ \Eprint {http://arxiv.org/abs/1707.04966}
  {arXiv:1707.04966} \BibitemShut {NoStop}%
%
\bibitem [{\citenamefont {Fukushima}\ \emph {et~al.}(2020)\citenamefont
  {Fukushima}, \citenamefont {Kojo},\ and\ \citenamefont
  {Weise}}]{Fukushima:2020cmk}%
  \BibitemOpen
  \bibfield  {author} {\bibinfo {author} {\bibfnamefont {K.}~\bibnamefont
  {Fukushima}}, \bibinfo {author} {\bibfnamefont {T.}~\bibnamefont {Kojo}}, \
  and\ \bibinfo {author} {\bibfnamefont {W.}~\bibnamefont {Weise}},\ }\href
  {\doibase 10.1103/PhysRevD.102.096017} {\bibfield  {journal} {\bibinfo
  {journal} {Phys. Rev. D}\ }\textbf {\bibinfo {volume} {102}},\ \bibinfo
  {pages} {096017} (\bibinfo {year} {2020})},\ \Eprint
  {http://arxiv.org/abs/2008.08436} {arXiv:2008.08436} \BibitemShut {NoStop}%
%
\bibitem [{\citenamefont {Glendenning}(2000)}]{Glendenning:2000}%
  \BibitemOpen
  \bibfield  {author} {\bibinfo {author} {\bibfnamefont {N.~K.}\ \bibnamefont
  {Glendenning}},\ }\href@noop {} {\emph {\bibinfo {title} {Compact Stars:
  {{Nuclear}} Physics, Particle Physics, and General Relativity}}}\ (\bibinfo
  {publisher} {{Springer}},\ \bibinfo {year} {2000})\BibitemShut {NoStop}%
%
\bibitem [{\citenamefont {Oertel}\ \emph {et~al.}(2017)\citenamefont {Oertel},
  \citenamefont {Hempel}, \citenamefont {Kl{\"a}hn},\ and\ \citenamefont
  {Typel}}]{Oertel:2016bki}%
  \BibitemOpen
  \bibfield  {author} {\bibinfo {author} {\bibfnamefont {M.}~\bibnamefont
  {Oertel}}, \bibinfo {author} {\bibfnamefont {M.}~\bibnamefont {Hempel}},
  \bibinfo {author} {\bibfnamefont {T.}~\bibnamefont {Kl{\"a}hn}}, \ and\
  \bibinfo {author} {\bibfnamefont {S.}~\bibnamefont {Typel}},\ }
  \href {\doibase 10.1103/RevModPhys.89.015007}  {\bibfield  {journal} {\bibinfo  {journal} {Rev. Mod. Phys.}\ }\textbf
  {\bibinfo {volume} {89}},\ \bibinfo {pages} {015007}  (\bibinfo {year} {2017})},\ \Eprint
  {http://arxiv.org/abs/1610.03361} {arXiv:1610.03361} \BibitemShut {NoStop}%
%
\bibitem [{\citenamefont {Lattimer}\ and\ \citenamefont
  {Prakash}(2004)}]{Lattimer:2004pg}%
  \BibitemOpen
  \bibfield  {author} {\bibinfo {author} {\bibfnamefont {J.~M.}\ \bibnamefont
  {Lattimer}}\ and\ \bibinfo {author} {\bibfnamefont {M.}~\bibnamefont
  {Prakash}},\ }\href {\doibase 10.1126/science.1090720} {\bibfield  {journal}
  {\bibinfo  {journal} {Science}\ }\textbf {\bibinfo {volume} {304}},\ \bibinfo
  {pages} {536} (\bibinfo {year} {2004})},\ \Eprint
  {http://arxiv.org/abs/astro-ph/0405262} {arXiv:astro-ph/0405262} \BibitemShut
  {NoStop}%
%
\bibitem [{\citenamefont {Weber}\ \emph {et~al.}(2007)\citenamefont {Weber},
  \citenamefont {Negreiros}, \citenamefont {Rosenfield},\ and\ \citenamefont
  {Stejner}}]{Weber:2006ep}%
  \BibitemOpen
  \bibfield  {author} {\bibinfo {author} {\bibfnamefont {F.}~\bibnamefont
  {Weber}}, \bibinfo {author} {\bibfnamefont {R.}~\bibnamefont {Negreiros}},
  \bibinfo {author} {\bibfnamefont {P.}~\bibnamefont {Rosenfield}}, \ and\
  \bibinfo {author} {\bibfnamefont {M.}~\bibnamefont {Stejner}},\ }\href
  {\doibase 10.1016/j.ppnp.2006.12.008} {\bibfield  {journal} {\bibinfo
  {journal} {Prog. Part. Nucl. Phys.}\ }\textbf {\bibinfo {volume} {59}},\
  \bibinfo {pages} {94} (\bibinfo {year} {2007})},\ \Eprint
  {http://arxiv.org/abs/astro-ph/0612054} {arXiv:astro-ph/0612054} \BibitemShut
  {NoStop}%
%
\bibitem [{\citenamefont {Demorest}\ \emph {et~al.}(2010)\citenamefont
  {Demorest}, \citenamefont {Pennucci}, \citenamefont {Ransom}, \citenamefont
  {Roberts},\ and\ \citenamefont {Hessels}}]{Demorest:2010bx}%
  \BibitemOpen
  \bibfield  {author} {\bibinfo {author} {\bibfnamefont {P.~B.}\ \bibnamefont
  {Demorest}}, \bibinfo {author} {\bibfnamefont {T.}~\bibnamefont {Pennucci}},
  \bibinfo {author} {\bibfnamefont {S.~M.}\ \bibnamefont {Ransom}}, \bibinfo
  {author} {\bibfnamefont {M.~S.~E.}\ \bibnamefont {Roberts}}, \ and\ \bibinfo
  {author} {\bibfnamefont {J.~W.~T.}\ \bibnamefont {Hessels}},\ }\href
  {\doibase 10.1038/nature09466} {\bibfield  {journal} {\bibinfo  {journal}
  {Nature}\ }\textbf {\bibinfo {volume} {467}},\ \bibinfo {pages} {1081}
  (\bibinfo {year} {2010})},\ \Eprint {http://arxiv.org/abs/1010.5788}
  {arXiv:1010.5788} \BibitemShut {NoStop}%
%
\bibitem [{\citenamefont {Antoniadis}\ \emph {et~al.}(2013)\citenamefont
  {Antoniadis} \emph {et~al.}}]{Antoniadis:2013pzd}%
  \BibitemOpen
  \bibfield  {author} {\bibinfo {author} {\bibfnamefont {J.}~\bibnamefont
  {Antoniadis}} \emph {et~al.},\ }\href {\doibase 10.1126/science.1233232}
  {\bibfield  {journal} {\bibinfo  {journal} {Science}\ }\textbf {\bibinfo
  {volume} {340}},\ \bibinfo {pages} {1233232} (\bibinfo {year} {2013})},\
  \Eprint {http://arxiv.org/abs/1304.6875} {arXiv:1304.6875} \BibitemShut
  {NoStop}%
%
\bibitem [{\citenamefont {Fonseca}\ \emph {et~al.}(2016)\citenamefont {Fonseca}
  \emph {et~al.}}]{Fonseca:2016tux}%
  \BibitemOpen
  \bibfield  {author} {\bibinfo {author} {\bibfnamefont {E.}~\bibnamefont
  {Fonseca}} \emph {et~al.},\ }\href {\doibase 10.3847/0004-637X/832/2/167}
  {\bibfield  {journal} {\bibinfo  {journal} {Astrophys. J.}\ }\textbf
  {\bibinfo {volume} {832}},\ \bibinfo {pages} {167} (\bibinfo {year}
  {2016})},\ \Eprint {http://arxiv.org/abs/1603.00545} {arXiv:1603.00545}
  \BibitemShut {NoStop}%
%
\bibitem [{\citenamefont {Arzoumanian}\ \emph {et~al.}(2018)\citenamefont
  {Arzoumanian} \emph {et~al.}}]{NANOGrav:2017wvv}%
  \BibitemOpen
  \bibfield  {author} {\bibinfo {author} {\bibfnamefont {Z.}~\bibnamefont
  {Arzoumanian}} \emph {et~al.} (\bibinfo {collaboration} {NANOGrav}),\ }\href
  {\doibase 10.3847/1538-4365/aab5b0} {\bibfield  {journal} {\bibinfo
  {journal} {Astrophys. J. Suppl.}\ }\textbf {\bibinfo {volume} {235}},\
  \bibinfo {pages} {37} (\bibinfo {year} {2018})}\BibitemShut {NoStop}%
%
\bibitem [{\citenamefont {Linares}\ \emph {et~al.}(2018)\citenamefont
  {Linares}, \citenamefont {Shahbaz},\ and\ \citenamefont
  {Casares}}]{Linares:2018ppq}%
  \BibitemOpen
  \bibfield  {author} {\bibinfo {author} {\bibfnamefont {M.}~\bibnamefont
  {Linares}}, \bibinfo {author} {\bibfnamefont {T.}~\bibnamefont {Shahbaz}}, \
  and\ \bibinfo {author} {\bibfnamefont {J.}~\bibnamefont {Casares}},\ }\href
  {\doibase 10.3847/1538-4357/aabde6} {\bibfield  {journal} {\bibinfo
  {journal} {Astrophys. J.}\ }\textbf {\bibinfo {volume} {859}},\ \bibinfo
  {pages} {54} (\bibinfo {year} {2018})},\ \Eprint
  {http://arxiv.org/abs/1805.08799} {arXiv:1805.08799} \BibitemShut {NoStop}%
%
\bibitem [{\citenamefont {Cromartie}\ \emph {et~al.}(2019)\citenamefont
  {Cromartie} \emph {et~al.}}]{NANOGrav:2019jur}%
  \BibitemOpen
  \bibfield  {author} {\bibinfo {author} {\bibfnamefont {H.~T.}\ \bibnamefont
  {Cromartie}} \emph {et~al.} (\bibinfo {collaboration} {NANOGrav}),\ }\href
  {\doibase 10.1038/s41550-019-0880-2} {\bibfield  {journal} {\bibinfo
  {journal} {Nature Astron.}\ }\textbf {\bibinfo {volume} {4}},\ \bibinfo
  {pages} {72} (\bibinfo {year} {2019})},\ \Eprint
  {http://arxiv.org/abs/1904.06759} {arXiv:1904.06759} \BibitemShut {NoStop}%
%
\bibitem [{\citenamefont {Fonseca}\ \emph {et~al.}(2021)\citenamefont {Fonseca}
  \emph {et~al.}}]{Fonseca:2021wxt}%
  \BibitemOpen
  \bibfield  {author} {\bibinfo {author} {\bibfnamefont {E.}~\bibnamefont
  {Fonseca}} \emph {et~al.},\ }\href {\doibase 10.3847/2041-8213/ac03b8}
  {\bibfield  {journal} {\bibinfo  {journal} {Astrophys. J. Lett.}\ }\textbf
  {\bibinfo {volume} {915}},\ \bibinfo {pages} {L12} (\bibinfo {year}
  {2021})},\ \Eprint {http://arxiv.org/abs/2104.00880} {arXiv:2104.00880}
  \BibitemShut {NoStop}%
%
\bibitem [{\citenamefont {Riley}\ \emph {et~al.}(2021)\citenamefont {Riley}
  \emph {et~al.}}]{Riley:2021pdl}%
  \BibitemOpen
  \bibfield  {author} {\bibinfo {author} {\bibfnamefont {T.~E.}\ \bibnamefont
  {Riley}} \emph {et~al.},\ }\href {\doibase 10.3847/2041-8213/ac0a81}
  {\bibfield  {journal} {\bibinfo  {journal} {Astrophys. J. Lett.}\ }\textbf
  {\bibinfo {volume} {918}},\ \bibinfo {pages} {L27} (\bibinfo {year}
  {2021})},\ \Eprint {http://arxiv.org/abs/2105.06980} {arXiv:2105.06980}
  \BibitemShut {NoStop}%
%
\bibitem [{\citenamefont {Miller}\ \emph {et~al.}(2021)\citenamefont {Miller}
  \emph {et~al.}}]{Miller:2021qha}%
  \BibitemOpen
  \bibfield  {author} {\bibinfo {author} {\bibfnamefont {M.~C.}\ \bibnamefont
  {Miller}} \emph {et~al.},\ }\href {\doibase 10.3847/2041-8213/ac089b}
  {\bibfield  {journal} {\bibinfo  {journal} {Astrophys. J. Lett.}\ }\textbf
  {\bibinfo {volume} {918}},\ \bibinfo {pages} {L28} (\bibinfo {year}
  {2021})},\ \Eprint {http://arxiv.org/abs/2105.06979} {arXiv:2105.06979}
  \BibitemShut {NoStop}%
%
\bibitem [{\citenamefont {Abbott}\ \emph
  {et~al.}(2017{\natexlab{a}})\citenamefont {Abbott} \emph
  {et~al.}}]{LIGOScientific:2017vwq}%
  \BibitemOpen
  \bibfield  {author} {\bibinfo {author} {\bibfnamefont {B.~P.}\ \bibnamefont
  {Abbott}} \emph {et~al.} (\bibinfo {collaboration} {LIGO Scientific,
  Virgo}),\ }\href {\doibase 10.1103/PhysRevLett.119.161101} {\bibfield
  {journal} {\bibinfo  {journal} {Phys. Rev. Lett.}\ }\textbf {\bibinfo
  {volume} {119}},\ \bibinfo {pages} {161101} (\bibinfo {year}
  {2017}{\natexlab{a}})},\ \Eprint {http://arxiv.org/abs/1710.05832}
  {arXiv:1710.05832} \BibitemShut {NoStop}%
%
\bibitem [{\citenamefont {Abbott}\ \emph
  {et~al.}(2017{\natexlab{b}})\citenamefont {Abbott} \emph
  {et~al.}}]{LIGOScientific:2017zic}%
  \BibitemOpen
  \bibfield  {author} {\bibinfo {author} {\bibfnamefont {B.~P.}\ \bibnamefont
  {Abbott}} \emph {et~al.} (\bibinfo {collaboration} {LIGO Scientific, Virgo,
  Fermi-GBM, INTEGRAL}),\ }\href {\doibase 10.3847/2041-8213/aa920c} {\bibfield
   {journal} {\bibinfo  {journal} {Astrophys. J. Lett.}\ }\textbf {\bibinfo
  {volume} {848}},\ \bibinfo {pages} {L13} (\bibinfo {year}
  {2017}{\natexlab{b}})},\ \Eprint {http://arxiv.org/abs/1710.05834}
  {arXiv:1710.05834} \BibitemShut {NoStop}%
%
\bibitem [{\citenamefont {Abbott}\ \emph {et~al.}(2020)\citenamefont {Abbott}
  \emph {et~al.}}]{LIGOScientific:2020aai}%
  \BibitemOpen
  \bibfield  {author} {\bibinfo {author} {\bibfnamefont {B.~P.}\ \bibnamefont
  {Abbott}} \emph {et~al.} (\bibinfo {collaboration} {LIGO Scientific,
  Virgo}),\ }\href {\doibase 10.3847/2041-8213/ab75f5} {\bibfield  {journal}
  {\bibinfo  {journal} {Astrophys. J. Lett.}\ }\textbf {\bibinfo {volume}
  {892}},\ \bibinfo {pages} {L3} (\bibinfo {year} {2020})},\ \Eprint
  {http://arxiv.org/abs/2001.01761} {arXiv:2001.01761} \BibitemShut {NoStop}%
%
\bibitem [{\citenamefont {Riley}\ \emph {et~al.}(2019)\citenamefont {Riley}
  \emph {et~al.}}]{Riley:2019yda}%
  \BibitemOpen
  \bibfield  {author} {\bibinfo {author} {\bibfnamefont {T.~E.}\ \bibnamefont
  {Riley}} \emph {et~al.},\ }\href {\doibase 10.3847/2041-8213/ab481c}
  {\bibfield  {journal} {\bibinfo  {journal} {Astrophys. J. Lett.}\ }\textbf
  {\bibinfo {volume} {887}},\ \bibinfo {pages} {L21} (\bibinfo {year}
  {2019})},\ \Eprint {http://arxiv.org/abs/1912.05702} {arXiv:1912.05702}
  \BibitemShut {NoStop}%
%
\bibitem [{\citenamefont {Miller}\ \emph {et~al.}(2019)\citenamefont {Miller}
  \emph {et~al.}}]{Miller:2019cac}%
  \BibitemOpen
  \bibfield  {author} {\bibinfo {author} {\bibfnamefont {M.~C.}\ \bibnamefont
  {Miller}} \emph {et~al.},\ }\href {\doibase 10.3847/2041-8213/ab50c5}
  {\bibfield  {journal} {\bibinfo  {journal} {Astrophys. J. Lett.}\ }\textbf {\bibinfo {volume} {887}},\ \bibinfo {pages} {L24}
  (\bibinfo {year} {2019})},\ \Eprint {http://arxiv.org/abs/1912.05705}
  {arXiv:1912.05705} \BibitemShut {NoStop}%
%
\bibitem [{\citenamefont {Annala}\ \emph {et~al.}(2018)\citenamefont {Annala},
  \citenamefont {Gorda}, \citenamefont {Kurkela},\ and\ \citenamefont
  {Vuorinen}}]{Annala:2017llu}%
  \BibitemOpen
  \bibfield  {author} {\bibinfo {author} {\bibfnamefont {E.}~\bibnamefont
  {Annala}}, \bibinfo {author} {\bibfnamefont {T.}~\bibnamefont {Gorda}},
  \bibinfo {author} {\bibfnamefont {A.}~\bibnamefont {Kurkela}}, \ and\
  \bibinfo {author} {\bibfnamefont {A.}~\bibnamefont {Vuorinen}},\ }\href
  {\doibase 10.1103/PhysRevLett.120.172703} {\bibfield  {journal} {\bibinfo
  {journal} {Phys. Rev. Lett.}\ }\textbf {\bibinfo {volume} {120}},\ \bibinfo
  {pages} {172703} (\bibinfo {year} {2018})},\ \Eprint
  {http://arxiv.org/abs/0912.0384} {arXiv:0912.0384} \BibitemShut {NoStop}%
%
\bibitem [{\citenamefont {Annala}\ \emph {et~al.}(2020)\citenamefont {Annala},
  \citenamefont {Gorda}, \citenamefont {Kurkela}, \citenamefont
  {N{\"a}ttil{\"a}},\ and\ \citenamefont {Vuorinen}}]{Annala:2019puf}%
  \BibitemOpen
  \bibfield  {author} {\bibinfo {author} {\bibfnamefont {E.}~\bibnamefont
  {Annala}}, \bibinfo {author} {\bibfnamefont {T.}~\bibnamefont {Gorda}},
  \bibinfo {author} {\bibfnamefont {A.}~\bibnamefont {Kurkela}}, \bibinfo
  {author} {\bibfnamefont {J.}~\bibnamefont {N{\"a}ttil{\"a}}}, \ and\ \bibinfo
  {author} {\bibfnamefont {A.}~\bibnamefont {Vuorinen}},\ }\href {\doibase
  10.1038/s41567-020-0914-9} {\bibfield  {journal} {\bibinfo  {journal} {Nature
  Phys.}\ }\textbf {\bibinfo {volume} {16}},\ \bibinfo {pages} {907} (\bibinfo
  {year} {2020})},\ \Eprint {http://arxiv.org/abs/1903.09121}
  {arXiv:1903.09121} \BibitemShut {NoStop}%
%
\bibitem [{\citenamefont {Masuda}\ \emph
  {et~al.}(2013{\natexlab{a}})\citenamefont {Masuda}, \citenamefont {Hatsuda},\
  and\ \citenamefont {Takatsuka}}]{Masuda:2012ed}%
  \BibitemOpen
  \bibfield  {author} {\bibinfo {author} {\bibfnamefont {K.}~\bibnamefont
  {Masuda}}, \bibinfo {author} {\bibfnamefont {T.}~\bibnamefont {Hatsuda}}, \
  and\ \bibinfo {author} {\bibfnamefont {T.}~\bibnamefont {Takatsuka}},\ }\href
  {\doibase 10.1093/ptep/ptt045} {\bibfield  {journal} {\bibinfo  {journal}
  {Prog. Theor. Exp. Phys.}\ }\textbf {\bibinfo {volume} {2013}},\ \bibinfo
  {pages} {073D01} (\bibinfo {year} {2013}{\natexlab{a}})},\ \Eprint
  {http://arxiv.org/abs/1212.6803} {arXiv:1212.6803} \BibitemShut {NoStop}%
%
\bibitem [{\citenamefont {Masuda}\ \emph
  {et~al.}(2013{\natexlab{b}})\citenamefont {Masuda}, \citenamefont {Hatsuda},\
  and\ \citenamefont {Takatsuka}}]{Masuda:2012kf}%
  \BibitemOpen
  \bibfield  {author} {\bibinfo {author} {\bibfnamefont {K.}~\bibnamefont
  {Masuda}}, \bibinfo {author} {\bibfnamefont {T.}~\bibnamefont {Hatsuda}}, \
  and\ \bibinfo {author} {\bibfnamefont {T.}~\bibnamefont {Takatsuka}},\ }\href
  {\doibase 10.1088/0004-637X/764/1/12} {\bibfield  {journal} {\bibinfo
  {journal} {Astrophys. J.}\ }\textbf {\bibinfo {volume} {764}},\ \bibinfo
  {pages} {12} (\bibinfo {year} {2013}{\natexlab{b}})},\ \Eprint
  {http://arxiv.org/abs/1205.3621} {arXiv:1205.3621} \BibitemShut {NoStop}%
%
\bibitem [{\citenamefont {Kojo}\ \emph {et~al.}(2015)\citenamefont {Kojo},
  \citenamefont {Powell}, \citenamefont {Song},\ and\ \citenamefont
  {Baym}}]{Kojo:2014rca}%
  \BibitemOpen
  \bibfield  {author} {\bibinfo {author} {\bibfnamefont {T.}~\bibnamefont
  {Kojo}}, \bibinfo {author} {\bibfnamefont {P.~D.}\ \bibnamefont {Powell}},
  \bibinfo {author} {\bibfnamefont {Y.~F.}\ \bibnamefont {Song}}, \ and\
  \bibinfo {author} {\bibfnamefont {G.}~\bibnamefont {Baym}},\ }\href {\doibase
  10.1103/PhysRevD.91.045003} {\bibfield  {journal} {\bibinfo  {journal} {Phys.
  Rev. D}\ }\textbf {\bibinfo {volume} {91}},\ \bibinfo {pages} {045003}
  (\bibinfo {year} {2015})},\ \Eprint {http://arxiv.org/abs/1412.1108}
  {arXiv:1412.1108} \BibitemShut {NoStop}%
%
\bibitem [{\citenamefont {Kojo}(2016)}]{Kojo:2015fua}%
  \BibitemOpen
  \bibfield  {author} {\bibinfo {author} {\bibfnamefont {T.}~\bibnamefont
  {Kojo}},\ }\href  {\doibase 10.1140/epja/i2016-16051-0} {\bibfield  {journal} {\bibinfo  {journal} {Eur.
  Phys. J. A}\ }\textbf {\bibinfo {volume} {52}},\ \bibinfo {pages} {51}  (\bibinfo {year} {2016})},\
  \Eprint {http://arxiv.org/abs/1508.04408} {arXiv:1508.04408} \BibitemShut  {NoStop}%
%
\bibitem [{\citenamefont {Wei}\ \emph {et~al.}(2020)\citenamefont {Wei},
  \citenamefont {Salinas}, \citenamefont {Kl{\"a}hn}, \citenamefont
  {Jaikumar},\ and\ \citenamefont {Barry}}]{Wei:2018tts}%
  \BibitemOpen
  \bibfield  {author} {\bibinfo {author} {\bibfnamefont {W.}~\bibnamefont
  {Wei}}, \bibinfo {author} {\bibfnamefont {M.}~\bibnamefont {Salinas}},
  \bibinfo {author} {\bibfnamefont {T.}~\bibnamefont {Kl{\"a}hn}}, \bibinfo
  {author} {\bibfnamefont {P.}~\bibnamefont {Jaikumar}}, \ and\ \bibinfo
  {author} {\bibfnamefont {M.}~\bibnamefont {Barry}},\ }\href {\doibase
  10.3847/1538-4357/abbe02} {\bibfield  {journal} {\bibinfo  {journal}
  {Astrophys. J.}\ }\textbf {\bibinfo {volume} {904}},\ \bibinfo {pages} {187}
  (\bibinfo {year} {2020})},\ \Eprint {http://arxiv.org/abs/1811.11377}
  {arXiv:1811.11377} \BibitemShut {NoStop}%
%
\bibitem [{\citenamefont {Jaikumar}\ \emph {et~al.}(2021)\citenamefont
  {Jaikumar}, \citenamefont {Semposki}, \citenamefont {Prakash},\ and\
  \citenamefont {Constantinou}}]{Jaikumar:2021jbw}%
  \BibitemOpen
  \bibfield  {author} {\bibinfo {author} {\bibfnamefont {P.}~\bibnamefont
  {Jaikumar}}, \bibinfo {author} {\bibfnamefont {A.}~\bibnamefont {Semposki}},
  \bibinfo {author} {\bibfnamefont {M.}~\bibnamefont {Prakash}}, \ and\
  \bibinfo {author} {\bibfnamefont {C.}~\bibnamefont {Constantinou}},\ }\href
  {\doibase 10.1103/PhysRevD.103.123009} {\bibfield  {journal} {\bibinfo
  {journal} {Phys. Rev. D}\ }\textbf {\bibinfo {volume} {103}},\ \bibinfo
  {pages} {123009} (\bibinfo {year} {2021})},\ \Eprint
  {http://arxiv.org/abs/2101.06349} {arXiv:2101.06349} \BibitemShut {NoStop}%
%
\bibitem [{\citenamefont {Bai}\ \emph {et~al.}(2021)\citenamefont {Bai},
  \citenamefont {Fu},\ and\ \citenamefont {Liu}}]{Bai:2021wrh}%
  \BibitemOpen
  \bibfield  {author} {\bibinfo {author} {\bibfnamefont {Z.}~\bibnamefont
  {Bai}}, \bibinfo {author} {\bibfnamefont {W.~J.}\ \bibnamefont {Fu}}, \ and\
  \bibinfo {author} {\bibfnamefont {Y.~X.}\ \bibnamefont {Liu}},\ }\href
  {\doibase 10.3847/1538-4357/ac2a31} {\bibfield  {journal} {\bibinfo
  {journal} {Astrophys. J.}\ }\textbf {\bibinfo {volume} {922}},\ \bibinfo
  {pages} {266} (\bibinfo {year} {2021})},\ \Eprint
  {http://arxiv.org/abs/2109.12614} {arXiv:2109.12614} \BibitemShut {NoStop}%
%
\bibitem [{\citenamefont {Tews}\ \emph {et~al.}(2018)\citenamefont {Tews},
  \citenamefont {Carlson}, \citenamefont {Gandolfi},\ and\ \citenamefont
  {Reddy}}]{Tews:2018kmu}%
  \BibitemOpen
  \bibfield  {author} {\bibinfo {author} {\bibfnamefont {I.}~\bibnamefont
  {Tews}}, \bibinfo {author} {\bibfnamefont {J.}~\bibnamefont {Carlson}},
  \bibinfo {author} {\bibfnamefont {S.}~\bibnamefont {Gandolfi}}, \ and\
  \bibinfo {author} {\bibfnamefont {S.}~\bibnamefont {Reddy}},\ }\href
  {\doibase 10.3847/1538-4357/aac267} {\bibfield  {journal} {\bibinfo
  {journal} {Astrophys. J.}\ }\textbf {\bibinfo {volume} {860}},\ \bibinfo
  {pages} {149} (\bibinfo {year} {2018})},\ \Eprint
  {http://arxiv.org/abs/1801.01923} {arXiv:1801.01923} \BibitemShut {NoStop}%
%
\bibitem [{\citenamefont {Alford}\ \emph {et~al.}(2013)\citenamefont {Alford},
  \citenamefont {Han},\ and\ \citenamefont {Prakash}}]{Alford:2013aca}%
  \BibitemOpen
  \bibfield  {author} {\bibinfo {author} {\bibfnamefont {M.~G.}\ \bibnamefont
  {Alford}}, \bibinfo {author} {\bibfnamefont {S.}~\bibnamefont {Han}}, \ and\
  \bibinfo {author} {\bibfnamefont {M.}~\bibnamefont {Prakash}},\ }\href
  {\doibase 10.1103/PhysRevD.88.083013} {\bibfield  {journal} {\bibinfo
  {journal} {Phys. Rev. D}\ }\textbf {\bibinfo {volume} {88}},\ \bibinfo
  {pages} {083013} (\bibinfo {year} {2013})},\ \Eprint
  {http://arxiv.org/abs/1302.4732} {arXiv:1302.4732} \BibitemShut {NoStop}%
%
\bibitem [{\citenamefont {Han}\ and\ \citenamefont
  {Steiner}(2019)}]{Han:2018mtj}%
  \BibitemOpen
  \bibfield  {author} {\bibinfo {author} {\bibfnamefont {S.}~\bibnamefont
  {Han}}\ and\ \bibinfo {author} {\bibfnamefont {A.~W.}\ \bibnamefont
  {Steiner}},\ }\href {\doibase 10.1103/PhysRevD.99.083014} {\bibfield
  {journal} {\bibinfo  {journal} {Phys. Rev. D}\ }\textbf {\bibinfo {volume}
  {99}},\ \bibinfo {pages} {083014} (\bibinfo {year} {2019})},\ \Eprint
  {http://arxiv.org/abs/1810.10967} {arXiv:1810.10967} \BibitemShut {NoStop}%
%
\bibitem [{\citenamefont {Li}\ \emph {et~al.}(2021)\citenamefont {Li},
  \citenamefont {Miao}, \citenamefont {Han},\ and\ \citenamefont
  {Zhang}}]{Li:2021crp}%
  \BibitemOpen
  \bibfield  {author} {\bibinfo {author} {\bibfnamefont {A.}~\bibnamefont
  {Li}}, \bibinfo {author} {\bibfnamefont {Z.~Q.}\ \bibnamefont {Miao}},
  \bibinfo {author} {\bibfnamefont {S.}~\bibnamefont {Han}}, \ and\ \bibinfo
  {author} {\bibfnamefont {B.}~\bibnamefont {Zhang}},\ }\href {\doibase
  10.3847/1538-4357/abf355} {\bibfield  {journal} {\bibinfo  {journal}
  {Astrophys. J.}\ }\textbf {\bibinfo {volume} {913}},\ \bibinfo {pages} {27}
  (\bibinfo {year} {2021})},\ \Eprint {http://arxiv.org/abs/2103.15119}
  {arXiv:2103.15119} \BibitemShut {NoStop}%
%
\bibitem [{\citenamefont {Greif}\ \emph {et~al.}(2019)\citenamefont {Greif},
  \citenamefont {Raaijmakers}, \citenamefont {Hebeler}, \citenamefont
  {Schwenk},\ and\ \citenamefont {Watts}}]{Greif:2018njt}%
  \BibitemOpen
  \bibfield  {author} {\bibinfo {author} {\bibfnamefont {S.~K.}\ \bibnamefont
  {Greif}}, \bibinfo {author} {\bibfnamefont {G.}~\bibnamefont {Raaijmakers}},
  \bibinfo {author} {\bibfnamefont {K.}~\bibnamefont {Hebeler}}, \bibinfo
  {author} {\bibfnamefont {A.}~\bibnamefont {Schwenk}}, \ and\ \bibinfo
  {author} {\bibfnamefont {A.~L.}\ \bibnamefont {Watts}},\ }\href {\doibase
  10.1093/mnras/stz654} {\bibfield  {journal} {\bibinfo  {journal} {Mon. Not.
  Roy. Astron. Soc.}\ }\textbf {\bibinfo {volume} {485}},\ \bibinfo {pages}
  {5363} (\bibinfo {year} {2019})},\ \Eprint {http://arxiv.org/abs/1812.08188}
  {arXiv:1812.08188} \BibitemShut {NoStop}%
%
\bibitem [{\citenamefont {Brandes}\ \emph {et~al.}(2022)\citenamefont
  {Brandes}, \citenamefont {Weise},\ and\ \citenamefont
  {Kaiser}}]{Brandes:2022nxa}%
  \BibitemOpen
  \bibfield  {author} {\bibinfo {author} {\bibfnamefont {L.}~\bibnamefont
  {Brandes}}, \bibinfo {author} {\bibfnamefont {W.}~\bibnamefont {Weise}}, \
  and\ \bibinfo {author} {\bibfnamefont {N.}~\bibnamefont {Kaiser}},\ }\href
  {\doibase 10.1103/PhysRevD.107.014011} {\bibfield  {journal} {\bibinfo
  {journal} {Phys. Rev. D}\ }\textbf {\bibinfo {volume} {107}},\ \bibinfo
  {pages} {014011} (\bibinfo {year} {2022})},\ \Eprint
  {http://arxiv.org/abs/2208.03026} {arXiv:2208.03026} \BibitemShut {NoStop}%
%
\bibitem [{\citenamefont {Tews}\ \emph {et~al.}(2019)\citenamefont {Tews},
  \citenamefont {Margueron},\ and\ \citenamefont {Reddy}}]{Tews:2019cap}%
  \BibitemOpen
  \bibfield  {author} {\bibinfo {author} {\bibfnamefont {I.}~\bibnamefont
  {Tews}}, \bibinfo {author} {\bibfnamefont {J.}~\bibnamefont {Margueron}}, \
  and\ \bibinfo {author} {\bibfnamefont {S.}~\bibnamefont {Reddy}},\ }\href
  {\doibase 10.1140/epja/i2019-12774-6} {\bibfield  {journal} {\bibinfo
  {journal} {Eur. Phys. J. A}\ }\textbf {\bibinfo {volume} {55}},\ \bibinfo
  {pages} {97} (\bibinfo {year} {2019})},\ \Eprint
  {http://arxiv.org/abs/1901.09874} {arXiv:1901.09874} \BibitemShut {NoStop}%
%
\bibitem [{\citenamefont {Han}\ \emph {et~al.}(2019)\citenamefont {Han},
  \citenamefont {Mamun}, \citenamefont {Lalit}, \citenamefont {Constantinou},\
  and\ \citenamefont {Prakash}}]{Han:2019bub}%
  \BibitemOpen
  \bibfield  {author} {\bibinfo {author} {\bibfnamefont {S.}~\bibnamefont
  {Han}}, \bibinfo {author} {\bibfnamefont {M.~A.~A.}\ \bibnamefont {Mamun}},
  \bibinfo {author} {\bibfnamefont {S.}~\bibnamefont {Lalit}}, \bibinfo
  {author} {\bibfnamefont {C.}~\bibnamefont {Constantinou}}, \ and\ \bibinfo
  {author} {\bibfnamefont {M.}~\bibnamefont {Prakash}},\ }\href {\doibase
  10.1103/PhysRevD.100.103022} {\bibfield  {journal} {\bibinfo  {journal}
  {Phys. Rev. D}\ }\textbf {\bibinfo {volume} {100}},\ \bibinfo {pages}
  {103022} (\bibinfo {year} {2019})},\ \Eprint
  {http://arxiv.org/abs/1906.04095} {arXiv:1906.04095} \BibitemShut {NoStop}%
%
\bibitem{Walecka:1974AP}
  \BibitemOpen
  \bibfield  {author} {\bibinfo {author} {\bibfnamefont {J. D. }~\bibnamefont
  {Walecka}}, \  }
  \href  {\doibase 10.1016/0003-4916(74)90208-5} {\bibfield  {journal} {\bibinfo  {journal} {Ann. Phys.}\
  }\textbf {\bibinfo {volume} {83}},\ \bibinfo {pages} {491} (\bibinfo {year} {1974})}.\
%
% `` A theory of highly condensed matter. "
%
\bibitem{Serot:1986ANP}
  \BibitemOpen
  \bibfield  {author} {\bibinfo {author} {\bibfnamefont {B. D.}~\bibnamefont
  {Serot}}\ and\ \bibinfo {author} {\bibfnamefont {J. D.}~\bibnamefont {Walecka}},\ }
  \href {\doibase } {\bibfield  {journal}
  {\bibinfo  {journal} {Adv. Nucl. Phys.}\ }\textbf {\bibinfo {volume} {16}},\
  \bibinfo {pages} {1} (\bibinfo {year} {1986}).}
%  \BibitemShut {NoStop}%
%
%  ``The Relativistic Nuclear Many Body Problem",
%         
\bibitem [{\citenamefont {McLerran}\ and\ \citenamefont
  {Pisarski}(2007)}]{McLerran:2007qj}%
  \BibitemOpen
  \bibfield  {author} {\bibinfo {author} {\bibfnamefont {L.}~\bibnamefont
  {McLerran}}\ and\ \bibinfo {author} {\bibfnamefont {R.~D.}\ \bibnamefont
  {Pisarski}},\ }\href {\doibase 10.1016/j.nuclphysa.2007.08.013} {\bibfield
  {journal} {\bibinfo  {journal} {Nucl. Phys. A}\ }\textbf {\bibinfo {volume}
  {796}},\ \bibinfo {pages} {83} (\bibinfo {year} {2007})},\ \Eprint
  {http://arxiv.org/abs/0706.2191} {arXiv:0706.2191} \BibitemShut {NoStop}%
%
\bibitem [{\citenamefont {Roberts}\ and\ \citenamefont
  {Williams}(1994)}]{Roberts:1994dr}%
  \BibitemOpen
  \bibfield  {author} {\bibinfo {author} {\bibfnamefont {C.~D.}\ \bibnamefont
  {Roberts}}\ and\ \bibinfo {author} {\bibfnamefont {A.~G.}\ \bibnamefont
  {Williams}},\ }\href {\doibase 10.1016/0146-6410(94)90049-3} {\bibfield
  {journal} {\bibinfo  {journal} {Prog. Part. Nucl. Phys.}\ }\textbf {\bibinfo
  {volume} {33}},\ \bibinfo {pages} {477} (\bibinfo {year} {1994})},\ \Eprint
  {http://arxiv.org/abs/hep-ph/9403224} {arXiv:hep-ph/9403224} \BibitemShut
  {NoStop}%
%
\bibitem [{\citenamefont {Maris}\ and\ \citenamefont
  {Roberts}(1997)}]{Maris:1997tm}%
  \BibitemOpen
  \bibfield  {author} {\bibinfo {author} {\bibfnamefont {P.}~\bibnamefont
  {Maris}}\ and\ \bibinfo {author} {\bibfnamefont {C.~D.}\ \bibnamefont
  {Roberts}},\ }\href {\doibase 10.1103/PhysRevC.56.3369} {\bibfield  {journal}
  {\bibinfo  {journal} {Phys. Rev. C}\ }\textbf {\bibinfo {volume} {56}},\
  \bibinfo {pages} {3369} (\bibinfo {year} {1997})},\ \Eprint
  {http://arxiv.org/abs/nucl-th/9708029} {arXiv:nucl-th/9708029} \BibitemShut
  {NoStop}%
%
\bibitem [{\citenamefont {Chang}\ and\ \citenamefont
  {Roberts}(2009)}]{Chang:2009zb}%
  \BibitemOpen
  \bibfield  {author} {\bibinfo {author} {\bibfnamefont {L.}~\bibnamefont
  {Chang}}\ and\ \bibinfo {author} {\bibfnamefont {C.~D.}\ \bibnamefont
  {Roberts}},\ }\href {\doibase 10.1103/PhysRevLett.103.081601} {\bibfield
  {journal} {\bibinfo  {journal} {Phys. Rev. Lett.}\ }\textbf {\bibinfo
  {volume} {103}},\ \bibinfo {pages} {081601} (\bibinfo {year} {2009})},\
  \Eprint {http://arxiv.org/abs/0903.5461} {arXiv:0903.5461} \BibitemShut
  {NoStop}%
%
\bibitem [{\citenamefont {Eichmann}\ \emph {et~al.}(2010)\citenamefont
  {Eichmann}, \citenamefont {Alkofer}, \citenamefont {Krassnigg},\ and\
  \citenamefont {Nicmorus}}]{Eichmann:2009qa}%
  \BibitemOpen
  \bibfield  {author} {\bibinfo {author} {\bibfnamefont {G.}~\bibnamefont
  {Eichmann}}, \bibinfo {author} {\bibfnamefont {R.}~\bibnamefont {Alkofer}},
  \bibinfo {author} {\bibfnamefont {A.}~\bibnamefont {Krassnigg}}, \ and\
  \bibinfo {author} {\bibfnamefont {D.}~\bibnamefont {Nicmorus}},\ }\href
  {\doibase 10.1103/PhysRevLett.104.201601} {\bibfield  {journal} {\bibinfo
  {journal} {Phys. Rev. Lett.}\ }\textbf {\bibinfo {volume} {104}},\ \bibinfo
  {pages} {201601} (\bibinfo {year} {2010})},\ \Eprint
  {http://arxiv.org/abs/0912.2246} {arXiv:0912.2246} \BibitemShut {NoStop}%
%
\bibitem{Ukawa:2015}
  \BibitemOpen
  \bibfield  {author} {\bibinfo {author} {\bibfnamefont {A.}~\bibnamefont
  {Ukawa}},\ }
  \href {\doibase 10.1007/s10955-015-1197-x} {\bibfield  {journal}
  {\bibinfo  {journal} {J. Stat. Phys.}\ }\textbf {\bibinfo {volume} {160}},\
  \bibinfo {pages} {1081} (\bibinfo {year} {2015})}\BibitemShut {NoStop}%
%
\bibitem [{\citenamefont {Typel}\ and\ \citenamefont
  {Wolter}(1999)}]{Typel:1999yq}%
  \BibitemOpen
  \bibfield  {author} {\bibinfo {author} {\bibfnamefont {S.}~\bibnamefont
  {Typel}}\ and\ \bibinfo {author} {\bibfnamefont {H.}~\bibnamefont {Wolter}},\
  }\href {\doibase 10.1016/S0375-9474(99)00310-3} {\bibfield  {journal}
  {\bibinfo  {journal} {Nucl. Phys. A}\ }\textbf {\bibinfo {volume} {656}},\
  \bibinfo {pages} {331} (\bibinfo {year} {1999})}\BibitemShut {NoStop}%
%
\bibitem [{\citenamefont {Chen}\ \emph {et~al.}(2011)\citenamefont {Chen},
  \citenamefont {Baldo}, \citenamefont {Burgio},\ and\ \citenamefont
  {Schulze}}]{Chen:2011my}%
  \BibitemOpen
  \bibfield  {author} {\bibinfo {author} {\bibfnamefont {H.}~\bibnamefont
  {Chen}}, \bibinfo {author} {\bibfnamefont {M.}~\bibnamefont {Baldo}},
  \bibinfo {author} {\bibfnamefont {G.~F.}\ \bibnamefont {Burgio}}, \ and\
  \bibinfo {author} {\bibfnamefont {H.~J.}\ \bibnamefont {Schulze}},\  }
  \href  {\doibase 10.1103/PhysRevD.84.105023} {\bibfield  {journal} {\bibinfo  {journal} {Phys. Rev. D}\
  }\textbf {\bibinfo {volume} {84}},\ \bibinfo {pages} {105023} (\bibinfo {year} {2011})},\
  \Eprint {http://arxiv.org/abs/1107.2497} {arXiv:1107.2497} \BibitemShut {NoStop}%
%
\bibitem [{\citenamefont {Bai}\ \emph {et~al.}(2018)\citenamefont {Bai},
  \citenamefont {Chen},\ and\ \citenamefont {Liu}}]{Bai:2017wvk}%
  \BibitemOpen
  \bibfield  {author} {\bibinfo {author} {\bibfnamefont {Z.}~\bibnamefont
  {Bai}}, \bibinfo {author} {\bibfnamefont {H.}~\bibnamefont {Chen}}, \ and\
  \bibinfo {author} {\bibfnamefont {Y.~X.}\ \bibnamefont {Liu}},\ }\href
  {\doibase 10.1103/PhysRevD.97.023018} {\bibfield  {journal} {\bibinfo
  {journal} {Phys. Rev. D}\ }\textbf {\bibinfo {volume} {97}},\ \bibinfo
  {pages} {023018} (\bibinfo {year} {2018})},\ \Eprint
  {http://arxiv.org/abs/1707.09535} {arXiv:1707.09535} \BibitemShut {NoStop}%
%
\bibitem [{\citenamefont {Wei}\ \emph {et~al.}(2019)\citenamefont {Wei},
  \citenamefont {Irving}, \citenamefont {Kl{\"a}hn},\ and\ \citenamefont
  {Jaikumar}}]{Wei:2018mxy}%
  \BibitemOpen
  \bibfield  {author} {\bibinfo {author} {\bibfnamefont {W.}~\bibnamefont
  {Wei}}, \bibinfo {author} {\bibfnamefont {B.}~\bibnamefont {Irving}},
  \bibinfo {author} {\bibfnamefont {T.}~\bibnamefont {Kl{\"a}hn}}, \ and\
  \bibinfo {author} {\bibfnamefont {P.}~\bibnamefont {Jaikumar}},\ }\href
  {\doibase 10.3847/1538-4357/ab53ea} {\bibfield  {journal} {\bibinfo
  {journal} {Astrophys. J.}\ }\textbf {\bibinfo {volume} {887}},\ \bibinfo
  {pages} {151} (\bibinfo {year} {2019})},\ \Eprint
  {http://arxiv.org/abs/1811.09441} {arXiv:1811.09441} \BibitemShut {NoStop}%
%
\bibitem [{\citenamefont {Husain}\ and\ \citenamefont
  {Thomas}(2021)}]{Husain:2020nbb}%
  \BibitemOpen
  \bibfield  {author} {\bibinfo {author} {\bibfnamefont {W.}~\bibnamefont
  {Husain}}\ and\ \bibinfo {author} {\bibfnamefont {A.~W.}\ \bibnamefont
  {Thomas}},\ }\href {\doibase 10.1063/5.0036994} {\bibfield  {journal}
  {\bibinfo  {journal} {AIP Conf. Proc.}\ }\textbf {\bibinfo {volume} {2319}},\
  \bibinfo {pages} {080001} (\bibinfo {year} {2021})},\ \Eprint
  {http://arxiv.org/abs/2010.06750} {arXiv:2010.06750} \BibitemShut {NoStop}%
%
\bibitem [{\citenamefont {Miyatsu}\ \emph {et~al.}(2015)\citenamefont
  {Miyatsu}, \citenamefont {Cheoun},\ and\ \citenamefont
  {Saito}}]{Miyatsu:2015kwa}%
  \BibitemOpen
  \bibfield  {author} {\bibinfo {author} {\bibfnamefont {T.}~\bibnamefont
  {Miyatsu}}, \bibinfo {author} {\bibfnamefont {M.-K.}\ \bibnamefont {Cheoun}},
  \ and\ \bibinfo {author} {\bibfnamefont {K.}~\bibnamefont {Saito}},\ }\href
  {\doibase 10.1088/0004-637X/813/2/135} {\bibfield  {journal} {\bibinfo
  {journal} {Astrophys. J.}\ }\textbf {\bibinfo {volume} {813}},\ \bibinfo
  {pages} {135} (\bibinfo {year} {2015})},\ \Eprint
  {http://arxiv.org/abs/1506.05552} {arXiv:1506.05552} \BibitemShut {NoStop}%
%
\bibitem [{\citenamefont {Flanagan}\ and\ \citenamefont
  {Hinderer}(2008)}]{Flanagan:2007ix}%
  \BibitemOpen
  \bibfield  {author} {\bibinfo {author} {\bibfnamefont {{\'E}.~{\'E}.}\
  \bibnamefont {Flanagan}}\ and\ \bibinfo {author} {\bibfnamefont
  {T.}~\bibnamefont {Hinderer}},\ }\href {\doibase 10.1103/PhysRevD.77.021502}
  {\bibfield  {journal} {\bibinfo  {journal} {Phys. Rev. D}\ }\textbf {\bibinfo
  {volume} {77}},\ \bibinfo {pages} {021502} (\bibinfo {year} {2008})},\
  \Eprint {http://arxiv.org/abs/0709.1915} {arXiv:0709.1915} \BibitemShut
  {NoStop}%
%
\bibitem [{\citenamefont {Zhao}\ and\ \citenamefont
  {Lattimer}(2018)}]{Zhao:2018nyf}%
  \BibitemOpen
  \bibfield  {author} {\bibinfo {author} {\bibfnamefont {T.~Q.}\ \bibnamefont
  {Zhao}}\ and\ \bibinfo {author} {\bibfnamefont {J.~M.}\ \bibnamefont
  {Lattimer}},\ }\href {\doibase 10.1103/PhysRevD.98.063020} {\bibfield
  {journal} {\bibinfo  {journal} {Phys. Rev. D}\ }\textbf {\bibinfo {volume}
  {98}},\ \bibinfo {pages} {063020} (\bibinfo {year} {2018})},\ \Eprint
  {http://arxiv.org/abs/1808.02858} {arXiv:1808.02858} \BibitemShut {NoStop}%
%
\bibitem [{\citenamefont {Margalit}\ and\ \citenamefont
  {Metzger}(2017)}]{Margalit:2017dij}%
  \BibitemOpen
  \bibfield  {author} {\bibinfo {author} {\bibfnamefont {B.}~\bibnamefont
  {Margalit}}\ and\ \bibinfo {author} {\bibfnamefont {B.~D.}\ \bibnamefont
  {Metzger}},\ }\href {\doibase 10.3847/2041-8213/aa991c} {\bibfield  {journal}
  {\bibinfo  {journal} {Astrophys. J. Lett.}\ }\textbf {\bibinfo {volume} {850}},\
  \bibinfo {pages} {L19} (\bibinfo {year} {2017})},\ \Eprint
  {http://arxiv.org/abs/1710.05938} {arXiv:1710.05938} \BibitemShut {NoStop}%
%
\bibitem [{\citenamefont {Rezzolla}\ \emph {et~al.}(2018)\citenamefont
  {Rezzolla}, \citenamefont {Most},\ and\ \citenamefont
  {Weih}}]{Rezzolla:2017aly}%
  \BibitemOpen
  \bibfield  {author} {\bibinfo {author} {\bibfnamefont {L.}~\bibnamefont
  {Rezzolla}}, \bibinfo {author} {\bibfnamefont {E.~R.}\ \bibnamefont {Most}},
  \ and\ \bibinfo {author} {\bibfnamefont {L.~R.}\ \bibnamefont {Weih}},\
  }\href {\doibase 10.3847/2041-8213/aaa401} {\bibfield  {journal} {\bibinfo
  {journal} {Astrophys. J. Lett.}\ }\textbf {\bibinfo {volume} {852}},\
  \bibinfo {pages} {L25} (\bibinfo {year} {2018})},\ \Eprint
  {http://arxiv.org/abs/1711.00314} {arXiv:1711.00314} \BibitemShut {NoStop}%
%
\bibitem [{\citenamefont {Ruiz}\ \emph {et~al.}(2018)\citenamefont {Ruiz},
  \citenamefont {Shapiro},\ and\ \citenamefont {Tsokaros}}]{Ruiz:2017due}%
  \BibitemOpen
  \bibfield  {author} {\bibinfo {author} {\bibfnamefont {M.}~\bibnamefont
  {Ruiz}}, \bibinfo {author} {\bibfnamefont {S.~L.}\ \bibnamefont {Shapiro}}, \
  and\ \bibinfo {author} {\bibfnamefont {A.}~\bibnamefont {Tsokaros}},\ }\href
  {\doibase 10.1103/PhysRevD.97.021501} {\bibfield  {journal} {\bibinfo
  {journal} {Phys. Rev. D}\ }\textbf {\bibinfo {volume} {97}},\ \bibinfo
  {pages} {021501} (\bibinfo {year} {2018})},\ \Eprint
  {http://arxiv.org/abs/1711.00473} {arXiv:1711.00473} \BibitemShut {NoStop}%
%
\bibitem [{\citenamefont {Raithel}\ and\ \citenamefont
  {Most}(2022)}]{Raithel:2022efm}%
  \BibitemOpen
  \bibfield  {author} {\bibinfo {author} {\bibfnamefont {C.~A.}\ \bibnamefont
  {Raithel}}\ and\ \bibinfo {author} {\bibfnamefont {E.~R.}\ \bibnamefont
  {Most}},\ }\href@noop {} {\  (\bibinfo {year} {2022})},\ \Eprint
  {http://arxiv.org/abs/2208.04294} {arXiv:2208.04294} \BibitemShut {NoStop}%
%
\bibitem [{\citenamefont {Liu}\ \emph {et~al.}(2001)\citenamefont {Liu},
  \citenamefont {Gao},\ and\ \citenamefont {Guo}}]{Liu:2001em}%
  \BibitemOpen
  \bibfield  {author} {\bibinfo {author} {\bibfnamefont {Y.~X.}\ \bibnamefont
  {Liu}}, \bibinfo {author} {\bibfnamefont {D.~F.}\ \bibnamefont {Gao}}, \ and\
  \bibinfo {author} {\bibfnamefont {H.}~\bibnamefont {Guo}},\ }\href {\doibase
  10.1016/S0375-9474(01)01120-4} {\bibfield  {journal} {\bibinfo  {journal}
  {Nucl. Phys. A}\ }\textbf {\bibinfo {volume} {695}},\ \bibinfo {pages} {353}
  (\bibinfo {year} {2001})},\ \Eprint {http://arxiv.org/abs/hep-ph/0105202}
  {arXiv:hep-ph/0105202} \BibitemShut {NoStop}%
%
\bibitem [{\citenamefont {Malik}\ \emph {et~al.}(2018)\citenamefont {Malik},
  \citenamefont {Alam}, \citenamefont {Fortin}, \citenamefont
  {Provid{\^e}ncia}, \citenamefont {Agrawal}, \citenamefont {Jha},
  \citenamefont {Kumar},\ and\ \citenamefont {Patra}}]{Malik:2018zcf}%
  \BibitemOpen
  \bibfield  {author} {\bibinfo {author} {\bibfnamefont {T.}~\bibnamefont
  {Malik}}, \bibinfo {author} {\bibfnamefont {N.}~\bibnamefont {Alam}},
  \bibinfo {author} {\bibfnamefont {M.}~\bibnamefont {Fortin}}, \bibinfo
  {author} {\bibfnamefont {C.}~\bibnamefont {Provid{\^e}ncia}}, \bibinfo
  {author} {\bibfnamefont {B.~K.}\ \bibnamefont {Agrawal}}, \bibinfo {author}
  {\bibfnamefont {T.~K.}\ \bibnamefont {Jha}}, \bibinfo {author} {\bibfnamefont
  {B.}~\bibnamefont {Kumar}}, \ and\ \bibinfo {author} {\bibfnamefont {S.~K.}\
  \bibnamefont {Patra}},\ }\href {\doibase 10.1103/PhysRevC.98.035804}
  {\bibfield  {journal} {\bibinfo  {journal} {Phys. Rev. C}\ }\textbf {\bibinfo
  {volume} {98}},\ \bibinfo {pages} {035804} (\bibinfo {year} {2018})},\
  \Eprint {http://arxiv.org/abs/1805.11963} {arXiv:1805.11963} \BibitemShut
  {NoStop}%
%
\bibitem [{\citenamefont {Lattimer}\ and\ \citenamefont
  {Steiner}(2014)}]{Lattimer:2014sga}%
  \BibitemOpen
  \bibfield  {author} {\bibinfo {author} {\bibfnamefont {J.~M.}\ \bibnamefont
  {Lattimer}}\ and\ \bibinfo {author} {\bibfnamefont {A.~W.}\ \bibnamefont
  {Steiner}},\ }\href {\doibase 10.1140/epja/i2014-14040-y} {\bibfield
  {journal} {\bibinfo  {journal} {Eur. Phys. J. A}\ }\textbf {\bibinfo {volume}
  {50}},\ \bibinfo {pages} {40} (\bibinfo {year} {2014})},\ \Eprint
  {http://arxiv.org/abs/1403.1186} {arXiv:1403.1186} \BibitemShut {NoStop}%
%
\bibitem [{\citenamefont {Abbott}\ \emph {et~al.}(2018)\citenamefont {Abbott}
  \emph {et~al.}}]{LIGOScientific:2018cki}%
  \BibitemOpen
  \bibfield  {author} {\bibinfo {author} {\bibfnamefont {B.~P.}\ \bibnamefont
  {Abbott}} \emph {et~al.} (\bibinfo {collaboration} {LIGO Scientific,
  Virgo}),\ }\href {\doibase 10.1103/PhysRevLett.121.161101} {\bibfield
  {journal} {\bibinfo  {journal} {Phys. Rev. Lett.}\ }\textbf {\bibinfo
  {volume} {121}},\ \bibinfo {pages} {161101} (\bibinfo {year} {2018})},\
  \Eprint {http://arxiv.org/abs/1805.11581} {arXiv:1805.11581} \BibitemShut
  {NoStop}%
%
\bibitem [{\citenamefont {Capano}\ \emph {et~al.}(2020)\citenamefont {Capano},
  \citenamefont {Tews}, \citenamefont {Brown}, \citenamefont {Margalit},
  \citenamefont {De}, \citenamefont {Kumar}, \citenamefont {Brown},
  \citenamefont {Krishnan},\ and\ \citenamefont {Reddy}}]{Capano:2019eae}%
  \BibitemOpen
  \bibfield  {author} {\bibinfo {author} {\bibfnamefont {C.~D.}\ \bibnamefont
  {Capano}}, \bibinfo {author} {\bibfnamefont {I.}~\bibnamefont {Tews}},
  \bibinfo {author} {\bibfnamefont {S.~M.}\ \bibnamefont {Brown}}, \bibinfo
  {author} {\bibfnamefont {B.}~\bibnamefont {Margalit}}, \bibinfo {author}
  {\bibfnamefont {S.}~\bibnamefont {De}}, \bibinfo {author} {\bibfnamefont
  {S.}~\bibnamefont {Kumar}}, \bibinfo {author} {\bibfnamefont {D.~A.}\
  \bibnamefont {Brown}}, \bibinfo {author} {\bibfnamefont {B.}~\bibnamefont
  {Krishnan}}, \ and\ \bibinfo {author} {\bibfnamefont {S.}~\bibnamefont
  {Reddy}},\ }\href {\doibase 10.1038/s41550-020-1014-6} {\bibfield  {journal}
  {\bibinfo  {journal} {Nature Astron.}\ }\textbf {\bibinfo {volume} {4}},\
  \bibinfo {pages} {625} (\bibinfo {year} {2020})},\ \Eprint
  {http://arxiv.org/abs/1908.10352} {arXiv:1908.10352} \BibitemShut {NoStop}%
\bibitem [{\citenamefont {Dutra}\ \emph {et~al.}(2016)\citenamefont {Dutra},
  \citenamefont {Louren{\c c}o},\ and\ \citenamefont
  {Menezes}}]{Dutra:2015hxa}%
  \BibitemOpen
  \bibfield  {author} {\bibinfo {author} {\bibfnamefont {M.}~\bibnamefont
  {Dutra}}, \bibinfo {author} {\bibfnamefont {O.}~\bibnamefont {Louren{\c
  c}o}}, \ and\ \bibinfo {author} {\bibfnamefont {D.~P.}\ \bibnamefont
  {Menezes}},\ }\href {\doibase 10.1103/PhysRevC.93.025806} {\bibfield
  {journal} {\bibinfo  {journal} {Phys. Rev. C}\ }\textbf {\bibinfo {volume}
  {93}},\ \bibinfo {pages} {025806}  (\bibinfo {year} {2016})},\ \Eprint
  {http://arxiv.org/abs/1510.02060} {arXiv:1510.02060} \BibitemShut {NoStop}%
%
\bibitem [{\citenamefont {Chang}\ \emph {et~al.}(2011)\citenamefont {Chang},
  \citenamefont {Liu},\ and\ \citenamefont {Roberts}}]{Chang:2010hb}%
  \BibitemOpen
  \bibfield  {author} {\bibinfo {author} {\bibfnamefont {L.}~\bibnamefont
  {Chang}}, \bibinfo {author} {\bibfnamefont {Y.~X.}\ \bibnamefont {Liu}}, \
  and\ \bibinfo {author} {\bibfnamefont {C.~D.}\ \bibnamefont {Roberts}},\
  }\href {\doibase 10.1103/PhysRevLett.106.072001} {\bibfield  {journal}
  {\bibinfo  {journal} {Phys. Rev. Lett.}\ }\textbf {\bibinfo {volume} {106}},\
  \bibinfo {pages} {072001} (\bibinfo {year} {2011})},\ \Eprint
  {http://arxiv.org/abs/1009.3458} {arXiv:1009.3458} \BibitemShut {NoStop}%
%
\bibitem [{\citenamefont {Qin}\ \emph {et~al.}(2011{\natexlab{b}})\citenamefont
  {Qin}, \citenamefont {Chang}, \citenamefont {Liu}, \citenamefont {Roberts},\
  and\ \citenamefont {Wilson}}]{Qin:2011dd}%
  \BibitemOpen
  \bibfield  {author} {\bibinfo {author} {\bibfnamefont {S.~X.}\ \bibnamefont
  {Qin}}, \bibinfo {author} {\bibfnamefont {L.}~\bibnamefont {Chang}}, \bibinfo
  {author} {\bibfnamefont {Y.~X.}\ \bibnamefont {Liu}}, \bibinfo {author}
  {\bibfnamefont {C.~D.}\ \bibnamefont {Roberts}}, \ and\ \bibinfo {author}
  {\bibfnamefont {D.~J.}\ \bibnamefont {Wilson}},\ }\href {\doibase
  10.1103/PhysRevC.84.042202} {\bibfield  {journal} {\bibinfo  {journal} {Phys.
  Rev. C}\ }\textbf {\bibinfo {volume} {84}},\ \bibinfo {pages} {042202}
  (\bibinfo {year} {2011}{\natexlab{b}})},\ \Eprint
  {http://arxiv.org/abs/1108.0603} {arXiv:1108.0603} \BibitemShut {NoStop}%
%
\bibitem [{\citenamefont {Chen}\ \emph {et~al.}(2008)\citenamefont {Chen},
  \citenamefont {Yuan}, \citenamefont {Chang}, \citenamefont {Liu},
  \citenamefont {Kl{\"a}hn},\ and\ \citenamefont {Roberts}}]{Chen:2008zr}%
  \BibitemOpen
  \bibfield  {author} {\bibinfo {author} {\bibfnamefont {H.}~\bibnamefont
  {Chen}}, \bibinfo {author} {\bibfnamefont {W.}~\bibnamefont {Yuan}}, \bibinfo
  {author} {\bibfnamefont {L.}~\bibnamefont {Chang}}, \bibinfo {author}
  {\bibfnamefont {Y.~X.}\ \bibnamefont {Liu}}, \bibinfo {author} {\bibfnamefont
  {T.}~\bibnamefont {Kl{\"a}hn}}, \ and\ \bibinfo {author} {\bibfnamefont
  {C.~D.}\ \bibnamefont {Roberts}},\ }\href {\doibase
  10.1103/PhysRevD.78.116015} {\bibfield  {journal} {\bibinfo  {journal} {Phys.
  Rev. D}\ }\textbf {\bibinfo {volume} {78}},\ \bibinfo {pages} {116015} (\bibinfo {year} {2008})},\
  \Eprint {http://arxiv.org/abs/0807.2755}  {arXiv:0807.2755} \BibitemShut {NoStop}%
%
\bibitem [{\citenamefont {Qin}\ \emph {et~al.}(2018)\citenamefont {Qin},
  \citenamefont {Roberts},\ and\ \citenamefont {Schmidt}}]{Qin:2018dqp}%
  \BibitemOpen
  \bibfield  {author} {\bibinfo {author} {\bibfnamefont {S.~X.}\ \bibnamefont
  {Qin}}, \bibinfo {author} {\bibfnamefont {C.~D.}\ \bibnamefont {Roberts}}, \
  and\ \bibinfo {author} {\bibfnamefont {S.~M.}\ \bibnamefont {Schmidt}},\
  }\href {\doibase 10.1103/PhysRevD.97.114017} {\bibfield  {journal} {\bibinfo
  {journal} {Phys. Rev. D}\ }\textbf {\bibinfo {volume} {97}},\ \bibinfo
  {pages} {114017} (\bibinfo {year} {2018})},\ \Eprint
  {http://arxiv.org/abs/1801.09697} {arXiv:1801.09697} \BibitemShut {NoStop}%
%
\bibitem [{\citenamefont {Qin}\ \emph {et~al.}(2019)\citenamefont {Qin},
  \citenamefont {Roberts},\ and\ \citenamefont {Schmidt}}]{Qin:2019hgk}%
  \BibitemOpen
  \bibfield  {author} {\bibinfo {author} {\bibfnamefont {S.~X.}\ \bibnamefont
  {Qin}}, \bibinfo {author} {\bibfnamefont {C.~D.}\ \bibnamefont {Roberts}}, \
  and\ \bibinfo {author} {\bibfnamefont {S.~M.}\ \bibnamefont {Schmidt}},\
  }\href {\doibase 10.1007/s00601-019-1488-x} {\bibfield  {journal} {\bibinfo
  {journal} {Few Body Syst.}\ }\textbf {\bibinfo {volume} {60}},\ \bibinfo
  {pages} {26} (\bibinfo {year} {2019})},\ \Eprint
  {http://arxiv.org/abs/1902.00026} {arXiv:1902.00026} \BibitemShut {NoStop}%
%
\bibitem [{\citenamefont {Alkofer}\ \emph {et~al.}(2002)\citenamefont
  {Alkofer}, \citenamefont {Watson},\ and\ \citenamefont
  {Weigel}}]{Alkofer:2002bp}%
  \BibitemOpen
  \bibfield  {author} {\bibinfo {author} {\bibfnamefont {R.}~\bibnamefont
  {Alkofer}}, \bibinfo {author} {\bibfnamefont {P.}~\bibnamefont {Watson}}, \
  and\ \bibinfo {author} {\bibfnamefont {H.}~\bibnamefont {Weigel}},\ }\href
  {\doibase 10.1103/PhysRevD.65.094026} {\bibfield  {journal} {\bibinfo
  {journal} {Phys. Rev. D}\ }\textbf {\bibinfo {volume} {65}},\ \bibinfo
  {pages} {094026} (\bibinfo {year} {2002})},\ \Eprint
  {http://arxiv.org/abs/hep-ph/0202053} {arXiv:hep-ph/0202053} \BibitemShut
  {NoStop}%
%
\bibitem [{\citenamefont {Chang}\ \emph {et~al.}(2005)\citenamefont {Chang},
  \citenamefont {Liu},\ and\ \citenamefont {Guo}}]{Chang:2005ay}%
  \BibitemOpen
  \bibfield  {author} {\bibinfo {author} {\bibfnamefont {L.}~\bibnamefont
  {Chang}}, \bibinfo {author} {\bibfnamefont {Y.~X.}\ \bibnamefont {Liu}}, \
  and\ \bibinfo {author} {\bibfnamefont {H.}~\bibnamefont {Guo}},\ }\href
  {\doibase 10.1016/j.nuclphysa.2005.01.020} {\bibfield  {journal} {\bibinfo
  {journal} {Nucl. Phys. A}\ }\textbf {\bibinfo {volume} {750}},\ \bibinfo
  {pages} {324} (\bibinfo {year} {2005})}\BibitemShut {NoStop}%
%
\bibitem [{\citenamefont {Chen}\ \emph {et~al.}(2012)\citenamefont {Chen},
  \citenamefont {Baldo}, \citenamefont {Burgio},\ and\ \citenamefont
  {Schulze}}]{Chen:2012zx}%
  \BibitemOpen
  \bibfield  {author} {\bibinfo {author} {\bibfnamefont {H.}~\bibnamefont
  {Chen}}, \bibinfo {author} {\bibfnamefont {M.}~\bibnamefont {Baldo}},
  \bibinfo {author} {\bibfnamefont {G.~F.}\ \bibnamefont {Burgio}}, \ and\
  \bibinfo {author} {\bibfnamefont {H.~J.}\ \bibnamefont {Schulze}},\ }
  \href {\doibase 10.1103/PhysRevD.86.045006}
  {\bibfield  {journal} {\bibinfo {journal} {Phys. Rev. D}\ }\textbf {\bibinfo {volume} {86}},\ \bibinfo {pages} {045006} (\bibinfo {year} {2012})},\
  \Eprint {http://arxiv.org/abs/1203.0158} {arXiv:1203.0158} \BibitemShut {NoStop}%
%
\bibitem [{\citenamefont {Chen}\ \emph {et~al.}(2016)\citenamefont {Chen},
  \citenamefont {Wei},\ and\ \citenamefont {Schulze}}]{Chen:2016ran}%
  \BibitemOpen
  \bibfield  {author} {\bibinfo {author} {\bibfnamefont {H.}~\bibnamefont
  {Chen}}, \bibinfo {author} {\bibfnamefont {J.~B.}\ \bibnamefont {Wei}}, \
  and\ \bibinfo {author} {\bibfnamefont {H.~J.}\ \bibnamefont {Schulze}},\
  }\href {\doibase 10.1140/epja/i2016-16291-x} {\bibfield  {journal} {\bibinfo
  {journal} {Eur. Phys. J. A}\ }\textbf {\bibinfo {volume} {52}},\ \bibinfo {pages} {291}
  (\bibinfo {year} {2016})},\
  \Eprint {http://arxiv.org/abs/1603.05755} {arXiv:1603.05755} \BibitemShut {NoStop}%
%
\bibitem [{\citenamefont {Chen}\ \emph {et~al.}(2015)\citenamefont {Chen},
  \citenamefont {Wei}, \citenamefont {Baldo}, \citenamefont {Burgio},\ and\
  \citenamefont {Schulze}}]{Chen:2015mda}%
  \BibitemOpen
  \bibfield  {author} {\bibinfo {author} {\bibfnamefont {H.}~\bibnamefont
  {Chen}}, \bibinfo {author} {\bibfnamefont {J.~B.}\ \bibnamefont {Wei}},
  \bibinfo {author} {\bibfnamefont {M.}~\bibnamefont {Baldo}}, \bibinfo
  {author} {\bibfnamefont {G.~F.}\ \bibnamefont {Burgio}}, \ and\ \bibinfo
  {author} {\bibfnamefont {H.~J.}\ \bibnamefont {Schulze}},\ }\href {\doibase
  10.1103/PhysRevD.91.105002} {\bibfield  {journal} {\bibinfo  {journal} {Phys.
  Rev. D}\ }\textbf {\bibinfo {volume} {91}},\ \bibinfo {pages} {105002}  (\bibinfo {year} {2015})},\
  \Eprint {http://arxiv.org/abs/1503.02795} {arXiv:1503.02795} \BibitemShut {NoStop}%
%
\bibitem [{\citenamefont {Kl{\"a}hn}\ \emph {et~al.}(2010)\citenamefont
  {Kl{\"a}hn}, \citenamefont {Roberts}, \citenamefont {Chang}, \citenamefont
  {Chen},\ and\ \citenamefont {Liu}}]{Klahn:2009mb}%
  \BibitemOpen
  \bibfield  {author} {\bibinfo {author} {\bibfnamefont {T.}~\bibnamefont
  {Kl{\"a}hn}}, \bibinfo {author} {\bibfnamefont {C.~D.}\ \bibnamefont
  {Roberts}}, \bibinfo {author} {\bibfnamefont {L.}~\bibnamefont {Chang}},
  \bibinfo {author} {\bibfnamefont {H.}~\bibnamefont {Chen}}, \ and\ \bibinfo
  {author} {\bibfnamefont {Y.~X.}\ \bibnamefont {Liu}},\ }\href {\doibase
  10.1103/PhysRevC.82.035801} {\bibfield  {journal} {\bibinfo  {journal} {Phys.
  Rev. C}\ }\textbf {\bibinfo {volume} {82}},\ \bibinfo {pages} {035801}  (\bibinfo {year} {2010})},\
  \Eprint {http://arxiv.org/abs/0911.0654} {arXiv:0911.0654} \BibitemShut {NoStop}
%
\bibitem [{\citenamefont {Haymaker}(1991)}]{Haymaker:1990vm}%
  \BibitemOpen
  \bibfield  {author} {\bibinfo {author} {\bibfnamefont {R.~W.}\ \bibnamefont
  {Haymaker}},\ }\href {\doibase 10.1007/BF02811226} {\bibfield  {journal}
  {\bibinfo  {journal} {Riv. Nuovo Cim.}\ }\textbf {\bibinfo {volume} {14}},\
  \bibinfo {pages} {1} (\bibinfo {year} {1991})}\BibitemShut {NoStop}%
\end{thebibliography}

%

\begin{appendix}

\section{Approaches Describing the Hadron Matter and the Quark Matter}\label{sec:Models}

In order to study the phase transition in compact star matter,
   one usually has to take approaches for hadron matter and quark matter separately.
We describe then the approaches we take in this work in this appendix.

\subsection{Relativistic Mean Field Theory for Hadron Matter}

For hadron matter, we adopt relativistic mean field (RMF) theory~\cite{Walecka:1974AP,Serot:1986ANP,Glendenning:2000,Typel:1999yq}.
The Lagrangian of the RMF reads:
\begin{equation}\label{eqn:Lagrangian}
\mathcal{L}=\mathcal{L}_{\mathrm{B}}^{} + \mathcal{L}_{\mathrm{lep}}^{} + \mathcal{L}_{\mathrm{M}}^{}
+ \mathcal{L}_{\mathrm{int}}^{} \, ,
\end{equation}
where $\mathcal{L}_{B}^{}$ is the Lagrangian of free baryons.

In this work, we consider the baryon octet $p$,$n$,$\Lambda$,$\Sigma^{\pm,0}$ and $\Xi^{-,0}$ for baryons.
The corresponding Lagrangian is written as
\begin{equation}
\mathcal{L}_{\mathrm{B}}^{} = \sum_{i}{\bar{\Psi}_{i}^{}} (\textrm{i}\gamma_{\mu}^{} \partial^{\mu} - m_{i}^{}){\Psi_{i}^{}}\, .
\end{equation}

$\mathcal{L}_{\mathrm{M}}$ is the Lagrangian of mesons, which reads
\begin{eqnarray}
\mathcal{L}_{\mathrm{M}}^{} &= & \frac{1}{2}\left(\partial_{\mu}^{} \sigma \partial^{\mu} \sigma - m_{\sigma}^{2} \sigma^{2} \right) -\frac{1}{4} \omega_{\mu\nu}^{} \omega^{\mu\nu}
- \frac{1}{2}m_{\omega}^{2} \omega_{\mu}\omega^{\mu}\nonumber\\
&& -\frac{1}{4}\boldsymbol{\rho}_{\mu\nu}^{} \boldsymbol{\rho}^{\mu\nu} -\frac{1}{2}m_{\rho}^{2} \boldsymbol{\rho}_{\mu} \boldsymbol{\rho}^{\mu} \, ,
\end{eqnarray}
where $\sigma$, $\omega_{\mu}^{}$, and $\boldsymbol{\rho}_{\mu}^{}$ are the
isoscalar-scalar, isoscalar-vector and isovector-vector meson field, respectively,
with $\omega_{\mu\nu}^{}=\partial_{\mu}\omega_{\nu} - \partial_{\nu}\omega_{\mu}$,
$\boldsymbol{\rho}_{\mu\nu}^{} =\partial_{\mu}\boldsymbol{\rho}_{\nu} - \partial_{\nu}\boldsymbol{\rho}_{\mu}$.

The $\mathcal{L}_{\mathrm{int}}$ in Eq.~(\ref{eqn:Lagrangian}) is the Lagrangian
 describing the interactions between baryons which are realized by exchanging the mesons:
\begin{eqnarray}
\mathcal{L}_{\mathrm{int}} = && \sum_{B} g_{\sigma B}^{} {\bar{\Psi}_{B}^{}} \sigma {\Psi_{B}^{}}
-g_{\omega B}^{} {\bar{\Psi}_{B}^{}} \gamma_{\mu}^{} \omega^{\mu} {\Psi_{B}^{}} \nonumber\\
		&&- g_{\rho B}^{} {\bar{\Psi}_{B}^{}} \gamma_{\mu}^{} \boldsymbol{\tau}_{B}^{} \cdot \boldsymbol{\rho}^{\mu} {\Psi_{B}^{}} \, ,
\end{eqnarray}
where $g_{iB}^{}$ for $i=\sigma$, $\omega$, $\rho$ are the coupling strength parameters between baryons and mesons, which depend on the baryon density.

For nucleons, the coupling constants are
%
%\begin{equation}
$$ {g_{iN}^{}}(\rho_{B}^{}) = {g_{iN}^{}} (\rho_{\mathrm{sat}}^{}) f_{i}^{} (x),  \qquad \quad \textrm{for} \quad i=\sigma,\omega, \rho \, , $$
%\end{equation}
%
where $\rho_{B}^{}$ is the baryon density, $\rho_{\mathrm{sat}}^{}$ is the saturation density of nuclear matter  and
$x= {\rho_{B}^{}}/{\rho_{\mathrm{sat}}^{}}$.
The density function can be written as~\cite{Typel:1999yq}:
\begin{eqnarray}
f_{i}^{}(x) & = & a_{i}^{} \frac{1+b_{i}^{} (x + d_{i}^{})^{2}}{1 + c_{i}^{}(x+d_{i}^{})^{2}}, \qquad \textrm{for} \qquad i=\sigma,\omega \, , \nonumber \\
f_{\rho}^{}(x)&  = &\textrm{exp}\left[-a_{\rho}^{} (x-1) \right] \, , \nonumber
\end{eqnarray}
where the value of parameters $a_{i}^{}$, $b_{i}^{}$, $c_{i}^{}$, $d_{i}^{}$ and $g_{iN}^{}(\rho_{\mathrm{sat}}^{})$
and $m_{i}^{}$ are listed in Table~\ref{tab:gi}.

\begin{table}[htb]
\caption{Parameters of the mesons and their couplings (taken from Ref.~\cite{Typel:1999yq}).}
\label{tab:gi}
\begin{tabular}{cccc}
\hline
Meson i & $\sigma$ & $\omega$ & $\rho$ \\
\hline
$m_i$(MeV) & 550 & 783 & 763\\
~$g_{iN}^{}(\rho_{\textrm{sat}}^{})$~ & ~$10.72854$~ & ~$13.29015$~ & ~$7.32196$~ \\
$a_{i}^{}$ & $1.365469$ & $1.402488$& $0.515$\\
$b_{i}^{}$ & $0.226061$ & $0.172577$& {} \\
$c_{i}^{}$ & $0.409704$ & $0.344293$& {} \\
$d_{i}^{}$ & $0.901995$ & $0.983955$& {} \\
\hline
\end{tabular}
\end{table}

For hyperons, we represent them with the relation between the hyperon coupling and the nucleon coupling as:
$\chi_{\sigma}^{}=\frac{g_{\sigma Y}^{}}{g_{\sigma N}^{}}$,
$\chi_{\omega}^{}=\frac{g_{\omega Y}^{}}{g_{\omega N}^{}}$,
$\chi_{\rho}^{}=\frac{g_{\rho Y}^{}}{g_{\rho N}^{}}$.
On the basis of hypernuclei experimental data, we choose them as those in Refs.~\cite{Glendenning:2000,Dutra:2015hxa}:
$\chi_{\sigma}^{}=0.7$, $\chi_{\omega}^{}=\chi_{\rho}^{}=0.783$.

The $\mathcal{L}_{\mathrm{lep}}^{}$ is the Lagrangian for leptons, which are treated as free fermions:
\begin{equation}
\mathcal{L}_{\mathrm{lep}}^{}=\sum_{l}^{} {\bar{\Psi}_{l}^{}}(\textrm{i} \gamma_{\mu}^{} \partial^{\mu} - m_{l}^{}) {\Psi_{l}^{}} \, ,
\end{equation}
and we include only the electron and muon in this work.

The field equations can be derived by differentiating the Lagrangian.
Under RMF approximation, the system is assumed to be in the static, uniform ground state.
The partial derivatives of the meson fields vanish,
except that the 0-component of the vector meson and the 3rd-component of the isovector meson survive and can be replaced with the corresponding expectation values.
The field equations of the mesons are then:
\begin{equation}\label{eqn:sigma}
m_{\sigma}^{2} \sigma =\sum_{B} g_{\sigma B}^{} \langle {\bar{\Psi}_{B}^{}} {\Psi_{B}^{}} \rangle \, ,
\end{equation}
\begin{equation}\label{eqn:omega}
m_{\omega}^{2} \omega_{0}^{} = \sum_{B} g_{\omega B}^{} \langle {\bar{\Psi}_{B}^{}} \gamma_{0}^{} {\Psi_{B}^{}} \rangle \, ,
\end{equation}
\begin{equation}\label{eqn:rho}
m_{\rho}^{2} \rho_{03}^{} = \sum_{B} g_{\rho B}^{} \langle {\bar{\Psi}_{B}^{}} \gamma_{0}^{} \tau_{3B}^{} {\Psi_{B}^{}} \rangle \, ,
\end{equation}
where $\tau_{3B}^{}$ is the 3rd-component of the isospin of baryon $B$.

The equation of motion (EoM) of the baryon is:
\begin{equation}\label{EOM}
\left[\gamma^{\mu} (\textrm{i}\partial_{\mu}^{} - \Sigma_{\mu}^{}) - (m_{B}^{} - g_{\sigma B}^{} \sigma) \right] {\Psi_{B}^{}} = 0 \, ,
\end{equation}
where
%\begin{equation}
$$ \Sigma_{\mu}^{} = g_{\omega B}^{} \omega_{\mu} + g_{\rho B}^{} \boldsymbol{\tau}_{B}^{} \cdot \boldsymbol{\rho}_{\mu}^{} + \Sigma_{\mu}^{\mathrm{R}} \, . $$
%\end{equation}
%
The $\Sigma_{\mu}^{\mathrm{R}}$ is called the ``rearrange'' term,
which appears because of the density-dependence of the coupling constant, and reads
\begin{eqnarray}\label{eqn:rearrange}
\Sigma_{\mu}^{\mathrm{R}} =&&  \frac{j_{\mu}^{}}{\rho} \bigg(\frac{\partial g_{\omega B}^{}} {\partial\rho}\bar{\Psi}_{B}^{} \gamma^{\nu} \Psi_{B}^{} \omega_{\nu}^{}
	 + \frac{\partial g_{\rho B}^{}}{\partial\rho}\bar{\Psi}_{B}^{} \gamma^{\nu} \boldsymbol{\tau}_{B}^{} \cdot \boldsymbol{\rho}_{\nu}^{} \Psi_{B}^{}\nonumber\\
&&		-\frac{\partial g_{\sigma B}^{}}{\partial\rho}\bar{\Psi}_{B}^{} \Psi_{B}^{} \sigma \bigg) \, ,
\end{eqnarray}
where $j_{\mu}^{} = \bar{\Psi}_{B}^{} \gamma_{\mu} \Psi_{B}^{}$ is the baryon current.

Under the EoM in Eq.~(\ref{EOM}),
the baryons behave as quasi-particles with effective mass
\begin{equation}\label{eqn:mstar}
m_{B}^{\ast} = m_{B}^{} - g_{\sigma B}^{} \sigma \, ,
\end{equation}
and effective chemical potential:
\begin{equation}\label{eqn:mustar}
\mu_{B}^{\ast} = \mu_{B}^{} - g_{\omega B}^{} \omega_{0}^{} - g_{\rho B}^{} \tau_{3B}^{} \rho_{03}^{} -\Sigma_{\mu}^{\textrm{R}} \, .
\end{equation}

One can then get the baryon (number) density:
\begin{equation}\label{eqn:ni}
\rho_{B}^{} \equiv \langle {\bar{\Psi}_{B}^{}} \gamma^{0} {\Psi_{B}^{}} \rangle =\gamma_{B}^{}\int\frac{\textrm{d}^3k}{(2\pi)^3} = \gamma_{B}^{}\frac{k_{FB}^{3}}{6\pi^{2}} \, ,
\end{equation}
where $k_{FB}^{} =\sqrt{\mu_{B}^{\ast 2} - m_{B}^{\ast 2}}$ is the Fermi momentum of the particle,
$\gamma_{B}^{} =2$ is the spin degeneracy.
And the scalar density is:
\begin{eqnarray}\label{eqn:nis}
	\rho_{B}^{s} &\equiv & \langle {\bar{\Psi}_{B}^{}} {\Psi_{B}^{}} \rangle =\gamma_{B}^{}\int\frac{\textrm{d}^3k}{(2\pi)^{3}}\frac{m_{B}^{\ast}}{\sqrt{k^{2} + m_{B}^{\ast 2}}}\nonumber\\
	&=&\gamma_{B}^{}\frac{m_{B}^{\ast}}{4\pi^{2}}\bigg[k_{FB}^{} \mu_{i}^{\ast} - m_{B}^{\ast 2} \textrm{ln}\bigg(\frac{k_{FB}^{} + \mu_{B}^{\ast}}{m_{B}^{\ast}}\bigg)\bigg] \, . \qquad
\end{eqnarray}

The density of a kind of leptons can be expressed as the same as that for baryons,
except that the effective mass and the effective chemical potential should be replaced with the corresponding mass and chemical potential of the lepton:
\begin{equation}\label{eqn:nl}
\rho_{l}^{} = \frac{k_{Fl}^{3}}{3\pi^{2}} \, ,
\end{equation}
where $k_{Fl}^{2} = \mu_{l}^{2} - m_{l}^{2}$ for $l=e^{-},\, \mu^{-}$.

The matter in the star composed of hadrons should be in $\beta$-equilibrium.
Since there are two conservation charge numbers: the baryon number and the electric charge number,
all the chemical potential can be expressed with the neutron chemical potential and the electron chemical potential:
\begin{equation}\label{eqn:beta}
\mu_{i}^{} = B \mu_{n}^{} - Q \mu_{e}^{} \, ,
\end{equation}
where $B$ and $Q$ is the baryon number, electric charge number for the particle $i$, respectively.

Then, combining Eqs.~(\ref{eqn:sigma}), (\ref{eqn:omega}), (\ref{eqn:rho}), (\ref{eqn:rearrange}), (\ref{eqn:mstar}),
(\ref{eqn:mustar}), (\ref{eqn:ni}), (\ref{eqn:nis}), (\ref{eqn:nl}) and (\ref{eqn:beta}),
together with the charge neutral condition:
%
%\begin{equation}
$$ \rho_{p}^{} + \rho_{\Sigma^{+}}^{}  = \rho_{e}^{} + \rho_{\mu^{-}}^{} + \rho_{\Sigma^{-}}^{}
+\rho_{\Xi^{-}}^{} \, , $$
%\end{equation}
%
one can determine the ingredients and the properties of the hadron matter with any given baryon density $\rho_{B}^{}$.

The EoS of the hadron matter can be calculated from the energy-momentum tensor:
\begin{equation}
T^{\mu\nu} = \sum_{\phi_i} \frac{\partial\mathcal{L}}{\partial(\partial_{\mu}\phi_{i}^{})}\partial^{\nu}\phi_{i}^{} - g^{\mu\nu}\mathcal{L}.
\end{equation}
The energy density $\varepsilon$ is:
\begin{equation}
\varepsilon=\langle T^{00}\rangle =\sum_{i=B,l}\varepsilon_{i}^{} +\frac{1}{2}m_{\sigma}^{2} \sigma^{2} +\frac{1}{2}m_{\omega}^{2} \omega_{0}^{2} + \frac{1}{2}m_{\rho}^{2} \rho_{03}^{2} \, ,
\end{equation}
where the contribution of the baryon $B$ to the energy density is:
\begin{equation}
\begin{split}
\varepsilon_{B}^{} & = \gamma_{B}^{}\int\frac{\textrm{d}^3k}{(2\pi)^3}\sqrt{k^{2} +m_{B}^{\ast 2}}\\
	=&  \frac{\gamma_{B}}{4\pi^{2}}\Big[ 2\mu_{B}^{\ast 3} k_{FB}^{}   -  m_{B}^{\ast 2} \mu_{B}^{\ast} k_{FB}^{}
 -  m_{B}^{\ast 4}\textrm{ln}\Big( \frac{\mu_{B}^{\ast} \! + \! k_{FB}^{}}{m_{B}^{\ast}}  \Big) \Big].
\end{split}
\end{equation}

The contribution of the leptons to the energy density can be written in the similar form as baryons with a spin degeneracy parameter $\gamma_{l}^{} = 2 $,
except that the effective mass and effective chemical potential should be replaced with those of the leptons, respectively.

As for the pressure of the system, we can determine that with the general formula:
\begin{equation}
P =\sum_{i} \mu_{i}^{} \rho_{i}^{} - \varepsilon \, .
\end{equation}

\subsection{Dyson-Schwinger Equation Approach for Quark Matter}

For quark matter, we adopt the Dyson-Schwinger (DS) equation approach of QCD (see, e.g., Refs.~\cite{Roberts:2000aa,Bashir:2012fs,Roberts:1994dr,Fischer:2018sdj,Chang:2010hb,Qin:2011dd}).
The DSE approach is a continuum field approach of QCD
which includes both the confinement and the dynamical chiral symmetry breaking features simultaneously~\cite{McLerran:2007qj},
and is successful in describing QCD phase transitions and hadron properties (see, e.g., Refs.~\cite{Roberts:1994dr,Roberts:2000aa,Eichmann:2009qa,Qin:2010nq,Bashir:2012fs,Chen:2008zr,Fischer:2018sdj,Gao:2015kea,Gao:2016qkh,Qin:2018dqp,Qin:2019hgk}).

The starting point of the DS equation approach is the stability of the quark, the gluon and the ghost fields.
The truncated one can be written as the quark gap equation:
\begin{equation}
S(p;\mu)^{-1}=Z_{2}^{} \big{[} \textrm{i}\boldsymbol{\gamma} \cdot  \boldsymbol{p} + \textrm{i}\gamma_{4}^{} (p_{4}^{} + \textrm{i}\mu) + m_{q}^{} \big{]} +\Sigma(p;\mu),
\end{equation}
where $S(p;\mu)$ is the quark propagator, $\Sigma(p;\mu)$ is the renormalized self-energy of the quark:
\begin{equation}
\begin{split}
\Sigma(p;\mu)=&Z_{1}^{} \int^{\Lambda}\frac{\textrm{d}^{4} q}{(2\pi)^{4}}
          g^{2}(\mu)D_{\rho\sigma}^{} (p-q;\mu)\\
		&\times\frac{\lambda^{a}}{2} \gamma_{\rho} S(q;\mu) \Gamma_{\sigma}^{a}(q,p;\mu),
\end{split}
\end{equation}
where $\int^{\Lambda}$ is the translationally regularized integral,
$\Lambda$ is the regularization mass-scale.
$g(\mu)$ is the strength of the coupling, $D_{\rho\sigma}^{}$ is the dressed gluon propagator,
$\Gamma_{\sigma}^{a}$ is the dressed quark-gluon interaction vertex,
$\lambda^{a}$ is the Gell-Mann matrix, and $m_{q}$ is the current mass of the quark.
For simplicity, the current mass of $u$ and $d$ quark is now taken to be zero,
and the current mass of $s$ quark to be $115\;$MeV, by fitting the kaon mass in vacuum~\cite{Chen:2011my}.
$Z_{1,2}^{}$ are the renormalization constants.

At finite chemical potential,
the quark propagator can be decomposed according to the Lorentz structure as:
\begin{equation}
\begin{split}
S(p;\mu)^{-1}=& \textrm{i} \boldsymbol{\gamma} \cdot  \boldsymbol{p} A(p^{2}, p \, u, \mu^{2})
+ B(p^{2},p\, u, \mu^{2})\\
		& + \textrm{i} \gamma_{4}^{} (p_{4} + \textrm{i} \mu) C(p^{2},p\, u, \mu^{2}) \, ,
\end{split}
\end{equation}
with $u=(\boldsymbol{0},\textrm{i}\mu)$.
A complete decomposition should include another term proportional to $\sigma_{\mu\nu}$,
but this term can be omitted in all practical calculations, since its contribution is extremely small.

At zero chemical potential, a commonly used ansatz for the dressed gluon propagator and the dressed quark-gluon interaction vertex is:
\begin{equation}
\begin{split}
 &Z_{1}^{} g^{2}  D_{\rho\sigma}^{}(p-q)\Gamma_{\sigma}^{a}(q,p)\\
		=&\mathcal{G}((p-q)^{2} )D_{\rho\sigma}^{\textrm{\small free}}(p-q) \frac{\lambda^a}{2}
             \Gamma_{\sigma}^{}(q,p) \, ,
\end{split}
\end{equation}
where
%\begin{equation}
$$ D_{\rho\sigma}^{\textrm{\small free}}(k\equiv p-q)=\frac{1}{k^{2}} \Big( \delta_{\rho\sigma}^{} - \frac{k_{\rho}^{} k_{\sigma}^{}}{k^{2}} \Big) \, , $$
%\end{equation}
%
$\mathcal{G}(k^{2})$ is the effective interaction introduced in the model,
and $\Gamma_{\sigma}^{}$ is the quark-gluon interaction vertex.

For the interaction part, we adopt the Gaussian type effective interaction (see, e.g., Refs.~\cite{Qin:2011dd,Chen:2008zr,Maris:1997tm,Chen:2011my,Alkofer:2002bp,Chang:2005ay}):
%
%\begin{equation}
$$ \frac{\mathcal{G}(k^{2})}{k^{2}}=\frac{4\pi^{2} D}{\omega^{6}}k^{2} \textrm{e}^{-k^{2}/\omega^{2}} \, , $$
%\end{equation}
%
where $D$ and $\omega$ are the parameters of the model.
In our present work we take $\omega=0.5\,$GeV and $D=1.0\,\textrm{ GeV}^2$ as the same as in
Refs.~\cite{Chang:2009zb,Chen:2011my,Chen:2012zx,Chen:2016ran,Bai:2017wvk}.
In case of finite chemical potential, an exponential dependence of the ${\mathcal{G}}$ on the chemical potential was introduced in Ref.~\cite{Chen:2011my} as:
\begin{equation}\label{eqn:alpha}
\frac{\mathcal{G}(k^{2};\mu)}{k^{2}} = \frac{4\pi^{2} D}{\omega^{6}} \textrm{e}^{-\alpha\mu^{2}/\omega^{2}}k^{2} \textrm{e}^{-k^{2}/\omega^{2}} \, ,
\end{equation}
where $\alpha$ is the parameter controlling the rate for the quark matter to approach the asymptotic freedom.
It is evident that, when $\alpha=0$, it is the same as that at zero chemical potential;
when $\alpha=\infty$, the effective interaction is zero and corresponds to the case of bag models.
We adopt such a model in our present calculations, and for simplicity, we take the same interaction for each flavor of the quarks.

For the quark--gluon interaction vertex,
Ref.~\cite{Chen:2015mda} has calculated the properties of quark matter with different vertex models
and shown that the vertex effect can be absorbed into the variation of the parameter $\alpha$.
We then in this work adopt only the rainbow approximation
%
%\begin{equation}
$ \Gamma_{\sigma}^{}(q,p) = \gamma_{\sigma}^{} \, . $\
%\end{equation}
%
In our previous work~\cite{Bai:2017wvk},
we have shown that when $\alpha=2$, the maximum mass of the hybrid star can be larger than $2M_{\odot}$.
We take then now $\alpha=2$.

With the above equations, we can get the quark propagator,
and derive the EoS of the quark matter in the same way as taken in Refs.~\cite{Qin:2010nq,Fischer:2018sdj,Chen:2008zr,Gao:2015kea,Gao:2016qkh,Klahn:2009mb}.

The number density of quarks as a function of its chemical potential is:
%\begin{equation}
$$ n_{q}^{}(\mu) = 6\int\frac{\textrm{d}^{3}\boldsymbol{p}}{(2\pi)^{3}} f_{q}^{}(|\boldsymbol{p}|;\mu) \, , $$
%\end{equation}
%
where $f_{q}^{}$ is the distribution function and reads
%\begin{equation}
$$ f_{q}^{}(|\boldsymbol{p}|;\mu) = \frac{1}{4\pi}\int_{-\infty}^{\infty}\textrm{d}p_{4}^{} \textrm{tr}_{\textrm{D}}^{}[-\gamma_{4}^{} S_{q}^{}(p;\mu)] \, , $$
%\end{equation}
%
where the trace is for the spinor indices.

The pressure of each flavor of quark at zero temperature can be obtained by integrating the number density:
\begin{equation}
P_{q}^{}(\mu_{q}^{}) = P_{q}^{}(\mu_{q,0}^{}) + \int_{\mu_{q,0}^{}}^{\mu_{q}^{}} n_{q}^{}(\mu) \textrm{d}\mu \, .
\end{equation}

The total pressure of the quark matter is the sum of the pressure of each flavor of quark:
\begin{equation}
P_{Q}^{}(\mu_{u}^{},\mu_{d}^{},\mu_{s}^{}) = \sum_{q=u,d,s}\tilde{P}_{q}^{}(\mu_{q}^{}) - B_{\mathrm{DS}}^{} \, ,
\end{equation}
\begin{equation}
\tilde{P}_{q}^{}(\mu_{q}^{})\equiv \int_{\mu_{q,0}^{}}^{\mu_{q}^{}} n_{q}^{}(\mu) \textrm{d}\mu \, ,
\end{equation}
\begin{equation}\label{eqn:B_DS}
B_{\mathrm{DS}}^{} \equiv -\sum_{q=u,d,s} P_{q}^{}(\mu_{q,0}^{}) \, .
\end{equation}

Theoretically, the starting point of the integral $\mu_{q,0}^{}$ can be any value,
in this work we take $\mu_{q,0}^{}=0$.
For the value of $B_{\mathrm{DS}}^{}$,
%
%along the line of the discussion in Ref.~\cite{Chen:2016EPJA},
%
we adopt the ``steepest-descent" approximation and implement $B_{\mathrm{DS}}^{} = 90\,\textrm{MeV}/{\textrm{fm}}^{3}$ as the same as taken in previous works~\cite{Chen:2008zr,Chen:2011my,Chen:2015mda,Haymaker:1990vm}.

The quark matter in a compact star should also be in $\beta$-equilibrium and electric charge neutral,
so we have:
%\begin{equation}
$$ \mu_{d}^{} = \mu_{u}^{} + \mu_{e}^{} = \mu_{s}^{} \, , $$
%\end{equation}
%
%\begin{equation}
$$ \frac{2\rho_{u}^{} - \rho_{d}^{} - \rho_{s}^{}}{3} - \rho_{e}^{} - \rho_{\mu^{-}}^{} = 0 \, . $$
%\end{equation}
%
And we have the baryon density and chemical potential as:
%
%\begin{equation}
$$ \rho_{B}^{} = \frac{1}{3}(\rho_{u}^{} + \rho_{d}^{} + \rho_{s}^{} ) \, , $$
%\end{equation}
%
\begin{equation}\label{muB}
\mu_{B}^{} = \mu_{u}^{} + 2\mu_{d}^{} \, .
\end{equation}

Therefore, we can calculate the properties of the quark matter with any given baryon chemical potential $\mu_{B}^{}$ or baryon density $\rho_{B}^{}$.

In this paper, we take $\rho_{Q}^{} = \rho_{B}^{}$ to denote the baryon density in quark matter.

\end{appendix}

\end{document}